\documentclass[aps,prd,nofootinbib,twocolumn,superscriptaddress,preprintnumbers,balancelastpage,longbibliography]{revtex4-1}

\usepackage{amsmath,amssymb,mathtools,bm}
\usepackage{graphicx, color, hepunits}
\usepackage[dvipsnames]{xcolor}
\usepackage{float}
\usepackage{filecontents}
\usepackage{multirow}
 \usepackage{hyperref}
\hypersetup{
    colorlinks=true,           linkcolor=blue,            citecolor=blue,            filecolor=magenta,         urlcolor=blue          }
\usepackage[utf8]{inputenc}
\usepackage[english]{babel}
\usepackage[nolist]{acronym}
\usepackage{xfrac}
\usepackage{xspace}

\let\vec\mathbf

\newcommand{\es}[2] {\begin{equation} \label{#1} \begin{split} #2 \end{split} \end{equation}}

\begin{document}

\title{The search for low-mass axion dark matter with ABRACADABRA-10\,cm}
\date{\today}
\author{Chiara~P.~Salemi}
\thanks{These authors contributed equally}
\email{salemi@mit.edu}
\affiliation{Laboratory of Nuclear Science, Massachusetts Institute of Technology, Cambridge, MA 02139}

\author{Joshua~W.~Foster}
\thanks{These authors contributed equally}
\email{fosterjw@umich.edu}
\affiliation{Leinweber Center for Theoretical Physics, Department of Physics, University of Michigan, Ann Arbor, MI 48109}
\affiliation{Berkeley Center for Theoretical Physics, University of California, Berkeley, CA 94720}
\affiliation{Theoretical Physics Group, Lawrence Berkeley National Laboratory, Berkeley, CA 94720}

\author{Jonathan~L.~Ouellet}
\thanks{These authors contributed equally}
\email{ouelletj@mit.edu}
\affiliation{Laboratory of Nuclear Science, Massachusetts Institute of Technology, Cambridge, MA 02139}

\author{Andrew~Gavin}
\affiliation{Department of Physics and Astronomy, University of North Carolina, Chapel Hill, Chapel Hill, NC, 27599}

\author{Kaliro\"e~M.~W.~Pappas}
\affiliation{Laboratory of Nuclear Science, Massachusetts Institute of Technology, Cambridge, MA 02139}

\author{Sabrina~Cheng}
\affiliation{Laboratory of Nuclear Science, Massachusetts Institute of Technology, Cambridge, MA 02139}

\author{Kate A. Richardson}
\affiliation{Department of Physics and Astronomy, University of North Carolina, Chapel Hill, Chapel Hill, NC, 27599}

\author{Reyco~Henning}
\affiliation{Department of Physics and Astronomy, University of North Carolina, Chapel Hill, Chapel Hill, NC, 27599}
\affiliation{Triangle Universities Nuclear Laboratory, Durham, NC 27710}

\author{Yonatan~Kahn}
\affiliation{Department of Physics, University of Illinois at Urbana-Champaign, Urbana, IL 61801}
\affiliation{Illinois Center for Advanced Studies of the Universe, University of Illinois at Urbana-Champaign, Urbana, IL 61801}

\author{Rachel~Nguyen}
\affiliation{Department of Physics, University of Illinois at Urbana-Champaign, Urbana, IL 61801}
\affiliation{Illinois Center for Advanced Studies of the Universe, University of Illinois at Urbana-Champaign, Urbana, IL 61801}

\author{Nicholas~L.~Rodd}
\affiliation{Berkeley Center for Theoretical Physics, University of California, Berkeley, CA 94720}
\affiliation{Theoretical Physics Group, Lawrence Berkeley National Laboratory, Berkeley, CA 94720}

\author{Benjamin~R.~Safdi}
\affiliation{Leinweber Center for Theoretical Physics, Department of Physics, University of Michigan, Ann Arbor, MI 48109}
\affiliation{Berkeley Center for Theoretical Physics, University of California, Berkeley, CA 94720}
\affiliation{Theoretical Physics Group, Lawrence Berkeley National Laboratory, Berkeley, CA 94720}

\author{Lindley~Winslow}
\email{lwinslow@mit.edu}
\affiliation{Laboratory of Nuclear Science, Massachusetts Institute of Technology, Cambridge, MA 02139}

\begin{abstract}
Two of the most pressing questions in physics are the microscopic nature of the dark matter that comprises 84\% of the mass in the universe and the absence of a neutron electric dipole moment.
These questions would be resolved by the existence of a hypothetical particle known as the quantum chromodynamics (QCD) axion.
In this work, we probe the hypothesis that axions constitute dark matter, using the \mbox{ABRACADABRA-10\,cm} experiment in a broadband configuration, with world-leading sensitivity.
We find no significant evidence for axions, and we present 95\% upper limits on the axion-photon coupling down to the world-leading level $g_{a\gamma\gamma}<3.2 \times10^{-11}$\,GeV$^{-1}$, representing one of the most sensitive searches for axions in the $0.41 - 8.27$\,neV mass range.
Our work paves a direct path for future experiments capable of confirming or excluding the hypothesis that dark matter is a QCD axion in the mass range motivated by String Theory and Grand Unified Theories.

\end{abstract}

\maketitle

\begin{acronym}
\acro{ADM}{axion dark datter}
\acro{SM}{Standard Model}
\acro{QCD}{quantum chromodynamics}
\acro{PQ}{Peccei-Quinn}
\acro{ALP}{axion-like particle}
\acro{WIMP}{Weakly Interacting Massive Particle}
\acro{PSD}{power spectral density}
\acro{DM}{dark matter}
\acro{DFT}{discrete Fourier transform}
\acro{FLL}{flux-lock feedback loop}
\acro{SNR}{signal-to-noise ratio}
\acro{LEE}{look-elsewhere effect}
\acro{TS}{test statistic}
\acro{POM}{polyoxymethylene}
\acro{PTFE}{polytetrafluoroethylene}
\acro{MC}{Monte Carlo}
\acro{AFS}{active feedback stabilization}
\acro{DR}{dilution refrigerator}
\end{acronym}

\newcommand{\ABRA}{\mbox{ABRACADABRA}\xspace}
\newcommand{\abra}{\mbox{ABRACADABRA-10\,cm}\xspace}
\newcommand{\ADM}{ADM\xspace}
\newcommand{\DM}{DM\xspace}
\newcommand{\ALP}{ALP\xspace}
\newcommand{\ALPs}{ALPs\xspace}
\newcommand{\WIMP}{WIMP\xspace}

\newcommand{\AFS}{AFS\xspace}
\newcommand{\PSD}{PSD\xspace}
\newcommand{\PSDs}{PSDs\xspace}
\newcommand{\DFT}{DFT\xspace}
\newcommand{\FLL}{FLL\xspace}
\newcommand{\SNR}{SNR\xspace}
\newcommand{\LEE}{LEE\xspace}
\newcommand{\TS}{TS\xspace}
\newcommand{\POM}{POM\xspace}
\newcommand{\PTFE}{PTFE\xspace}
\newcommand{\MC}{MC\xspace}
\newcommand{\DR}{DR\xspace}

\newcommand{\rhoDM}{\ensuremath{\rho_{\rm DM}}\xspace}
\newcommand{\gagg}{\ensuremath{g_{a\gamma\gamma}}\xspace}
\newcommand{\Jeff}{\ensuremath{\mathbf{J}_{\rm eff}}\xspace}

\newcommand{\Px}{\ensuremath{\bar{\mathcal{F}}}}
\newcommand{\Pten}{\ensuremath{\bar{\mathcal{F}}_\mathrm{10M}}\xspace}
\newcommand{\Pone}{\ensuremath{\bar{\mathcal{F}}_\mathrm{1M}}\xspace}
\newcommand{\Phun}{\ensuremath{\mathcal{F}_\mathrm{100k}}\xspace}

The axion is a well-motivated candidate to explain the particle nature of dark matter (\DM)~\cite{Preskill:1982cy,Abbott:1982af,Dine:1982ah}. This pseudoscalar particle is naturally realized as a pseudo-Goldstone boson of the Peccei-Quinn (PQ) symmetry, which is broken at a high scale $f_a$; the axion would be exactly massless but for its low-energy interactions with quantum chromodynamics (QCD)~\cite{Peccei:1977ur,Peccei:1977hh,Weinberg:1977ma,Wilczek:1977pj}. The axion mass is tied to the scale $f_a$ by \mbox{$m_a \approx 5.7 (10^{15} \, \, {\rm GeV} / f_a)$ neV}~\cite{Borsanyi:2016ksw}.  The range of scales $f_a \approx 10^{15}-10^{16}$\,GeV is particularly compelling because of connections to String Theory~\cite{Svrcek:2006yi} and Grand Unification~\cite{DiLuzio:2018gqe,Co:2016vsi}, and in the corresponding mass range of $m_a \sim 1-10 \ {\rm neV}$ the axion may naturally explain the observed \DM abundance~\cite{Tegmark:2005dy,Co:2016vsi}.  In this Article we provide the most sensitive probe of axion dark datter (\ADM) in this mass range to date. 

\ADM that couples to photons modifies Amp\`ere's law such that in current-free regions
\begin{equation}
    \vec{\nabla}\times\textbf{B}=\frac{\partial\textbf{E}}{\partial t}-g_{a\gamma\gamma}\left(\textbf{E}\times\vec{\nabla}a-\frac{\partial a}{\partial t}\textbf{B}\right) \,,
\end{equation}
with ${\bf E}$ and ${\bf B}$ the electric and magnetic fields, respectively, $a(\textbf{x},t)$ the \ADM field, and $g_{a\gamma\gamma}$ the axion-electromagnetic coupling constant.  In the presence of a static external magnetic field \ADM behaves like an effective current density $\Jeff = g_{a\gamma\gamma} ( \partial_t a ) {\bf B}$. If the axion makes up all of the observed \DM then, to leading order in the DM velocity, $\partial_t a  = \sqrt{2 \rhoDM} \cos( m_a t)$, with $\rhoDM \approx 0.4\,\mathrm{GeV/cm}^3$ the local \DM density \cite{Read:2014qva}.  It was pointed out in~\cite{Sikivie:2013laa,ABRA2016} that the effective current induces an oscillating secondary magnetic field which may be detectable in the laboratory without the aid of a resonant cavity for sufficiently small $m_a$.  The oscillation frequency is given by $f = m_a / (2 \pi)$, with bandwidth $\delta f / f \approx 10^{-6}$ arising from the finite axion velocity dispersion~\cite{Sikivie:1983ip}.
In this work we leverage this theoretical principle to search for axions in the laboratory. 

The most common detection strategy for \ADM is through the electromagnetic coupling $\gagg$, which for the QCD axion is directly proportional to the mass $m_a$.
Until recently, experiments have focused on searching for axions in the mass range $1\lesssim m_a \lesssim 40\,\mu\mathrm{eV}$, which is well-suited to microwave cavity searches \cite{PhysRevD.42.1297, Asztalos2001,ADMX2018,HAYSTAC2018a,Backes:2020ajv}. In the low-mass regime targeted in this work, the Compton wavelength of the axion $\lambda_C \sim {\rm km}$ is much larger than the experimental apparatus, and so the sensitivity of the experiment improves with volume as $V^{5/6}$, roughly independent of $m_a$ until the size of the experiment approaches $\lambda_C$ \cite{ABRA2016}. This scaling is important because the expected coupling $\gagg$ is smaller at lower masses, requiring ever-more-sensitive experiments to 
achieve a detection. 
\ABRA is an experimental program designed to detect axions at the Grand Unification scale using a strong toroidal magnetic field~\cite{ABRA2016}.  \ABRA is part of a suite of ADM experiments which together aim to probe the full QCD axion parameter space~\cite{JacksonKimball:2017elr,McAllister:2017lkb,Silva-Feaver2016,DMRadio_Design,ADMX2018,HAYSTAC2018a,Backes:2020ajv,TheMADMAXWorkingGroup:2016hpc,Gramolin2020a,Lee:2020cfj}.
The experiment we report on here, \abra, is a prototype for a larger \ADM detector that would be sensitive to the QCD axion.
This Article presents data collected in 2020 that is up to an order of magnitude more sensitive than our previous results~\cite{Ouellet:2018beu} and places strong limits on \ADM in the {$0.41 - 8.27$\,neV} range of axion masses.

\section*{ABRACADABRA-10\,cm detector}

\begin{figure}
    \centering
    \includegraphics[width=.8\columnwidth]{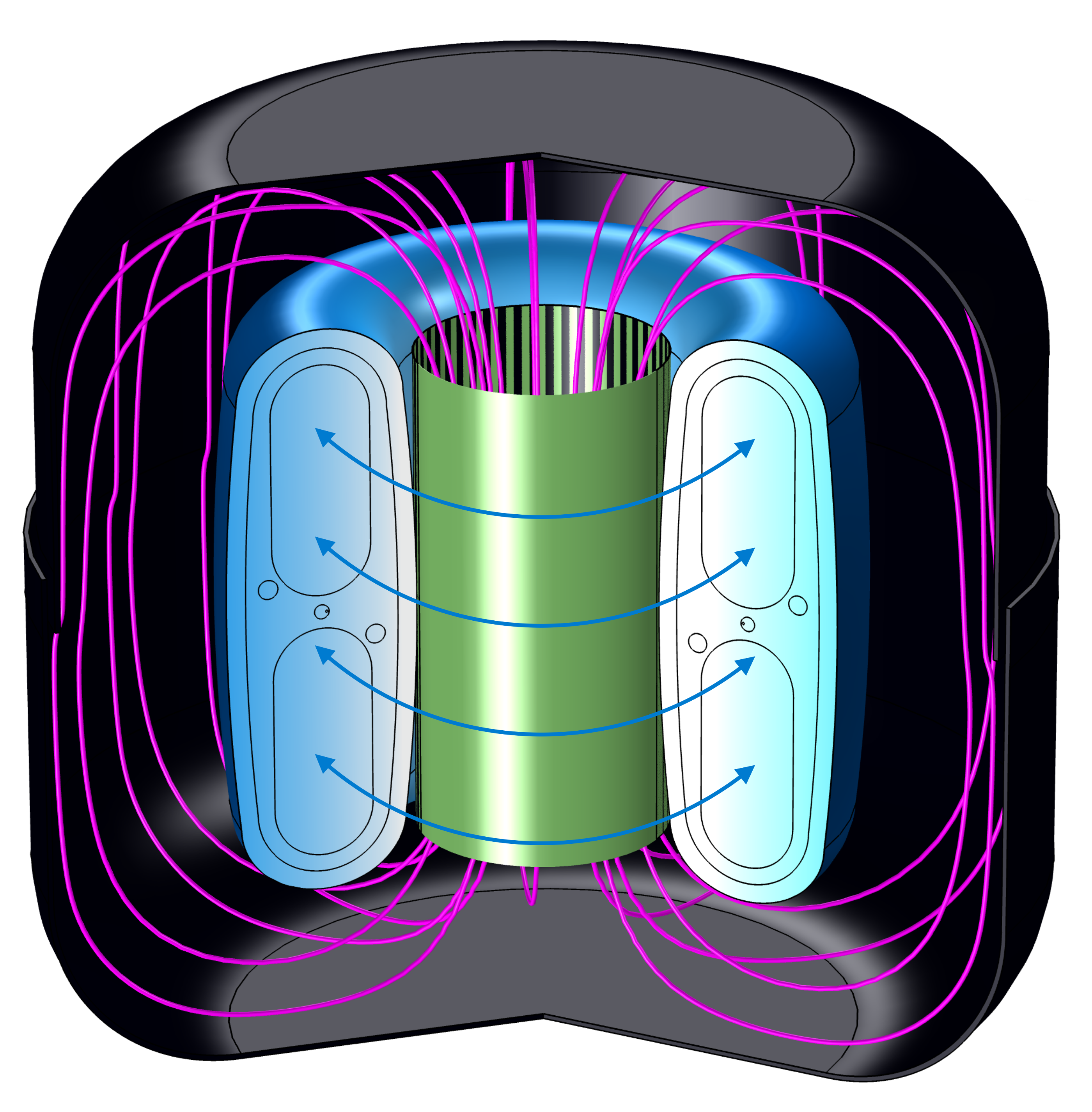}\\
    \includegraphics[width=\columnwidth]{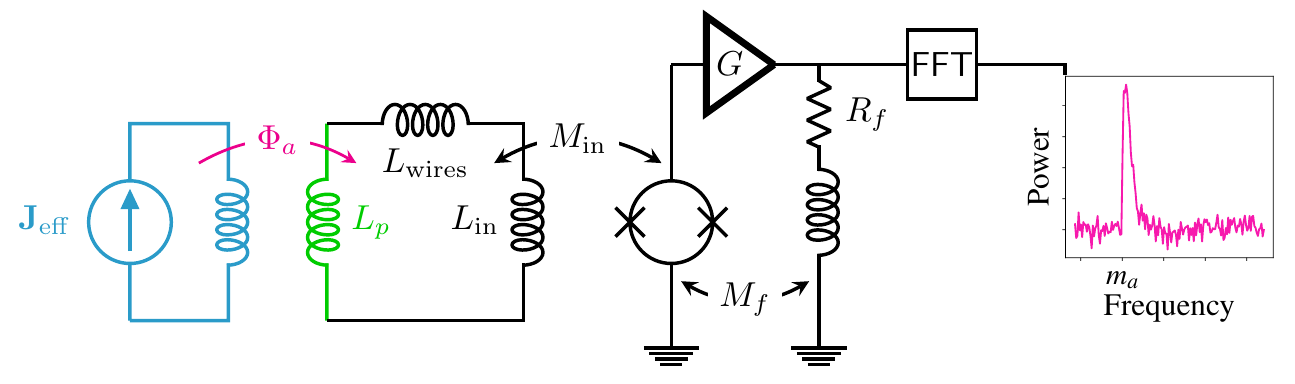}
    \caption{
    \emph{Top}: Schematic of \abra showing the effective axion-induced current (blue), sourced by the toroidal magnetic field, generating a magnetic flux (magenta) through the pickup cylinder (green) in the toroid bore. \emph{Bottom}: Simplified schematic of the \abra readout (full circuit diagram in Supp. Fig.~\ref{fig:circuits}). The pickup cylinder $L_p$ is inductively coupled to the axion effective current \Jeff. The power spectrum of the induced current is read out through a DC SQUID inductively coupled to the circuit through $L_{\rm in}$. An axion signal would appear as excess power above the noise floor at a frequency corresponding to the axion mass.}
    \label{fig:detector}
\end{figure}

The ABRACADABRA-10\,cm detector is built around a 12\,cm diameter, 12\,cm tall, 1\,T toroidal magnet fabricated by Superconducting Systems Inc \cite{ssi}.  The axion interactions with the toroidal magnetic field $\mathbf{B}_0$ drive the effective current, \Jeff, which oscillates parallel to $\mathbf{B}_0$ and sources a real oscillating magnetic field through the toroid's center.  The oscillating magnetic flux is read out with a two-stage DC-SQUID  via a superconducting pickup in the central bore.   Unlike other axion detector designs, this novel geometry situates the readout pickup in a nominally field-free region unless axions are present  \cite{ABRA2016}.  The detector can be calibrated by injecting fake axion signals ({\it i.e.},\ AC currents) through a wire calibration loop that runs through the body of the magnet.  The detector, illustrated schematically in Fig.\,\ref{fig:detector}, is located on MIT's campus in Cambridge, MA.

In 2019, we performed several detector upgrades from the Run~1 configuration in order to improve our sensitivity~\cite{Ouellet:2018beu,ABRA_10cm_Technical}. 
In this Article we report the results of the subsequent data campaign (Run~3), collected after the detector upgrade.  Run~3 data consists of $\sim$430 hours of data collected from June 5 to June 29, 2020. 

Before the upgrades were complete, we took additional, uncalibrated data (Run~2), which is not presented here.  A subset of that data was instead used to develop our data analysis procedure in order to run a blind analysis on the Run~3 data, as described in detail below.  

The total expected axion power, $A$,  coupled into our readout pickup is related to the axion-induced flux $\Phi_a$ as 
\begin{equation}
A\equiv \langle|\Phi_a|^2 \rangle = \gagg^2\rhoDM \mathcal{G}^2V^2B^2_{\rm max},
\label{eqn:coupled_power}
\end{equation}
where $\mathcal{G}$ is a geometric coupling, $V$ is the magnetic field volume, $B_{\rm max}$ is the maximum value of $|\mathbf{B}_0|$, and the angle brackets denote the time average~\cite{ABRA2016,Foster2018}.
Run~1 utilized a 4.02\,cm diameter pickup loop made from a 1\,mm diameter wire, giving $\mathcal{G}\approx0.027$. In 2019, we replaced this readout with a 10\,cm tall, 5.1\,cm diameter superconducting cylinder pickup centered in the toroid bore. This consisted of a 150~$\mu$m-thick Nb sheet wrapped around a polytetrafluoroethylene (PTFE) cylinder. This design yields a stronger geometric coupling to \Jeff of $\mathcal{G}\approx0.031$ and decreases the inductance of the pickup \cite{ABRA2016}. We compute $\mathcal{G}$ using electromagnetic simulations in the COMSOL Multiphysics package \cite{ABRA_10cm_Technical,COMSOL}.

To amplify our signal, $\Phi_a$ is coupled into the readout SQUID through the pickup circuit (see Fig.~\ref{fig:detector}) yielding a transformer gain $M_{\rm in}/L_T$, where $M_{\rm in}$ is the input coupling to the SQUID, and $L_T\equiv L_p+L_{\rm in} + L_{\rm wires}$ is the total inductance of the pickup circuit, with $L_p$ the pickup cylinder inductance, $L_{\rm in}$ the input inductance of the SQUID package, and $L_{\rm wires}$ the parasitic inductance, dominated by the twisted pair wiring. The SQUID, manufactured by Magnicon~\cite{magnicon}, is read out using Magnicon's XXF-1 SQUID electronics operating in closed feedback loop mode. The Run~1 sensitivity was limited by parasitic inductance in the NbTi wiring of this circuit that placed a lower limit on $L_T \gtrsim1.6\,\mu$H. During the upgrade, we replaced this wiring, moving the SQUIDs closer to the detector to reduce the wire length.  Based on calibration data, we found that the total impedance in the circuit is $\sim800\,$nH.  Finally, the SQUID was operated at a higher flux-to-voltage gain setting of 4.3\,V/$\Phi_0$ in Run~3, compared to the previous Run~1 which we ran at 1.29\,V/$\Phi_0$ due to higher levels of environmental noise.  This change does not directly improve the signal gain, but does reduce system noise.  We also improved our noise floor by reducing the operating temperature of the SQUID package from $\sim$870\,mK to $\sim$450\,mK.
All together, the upgrade campaign increased the expected power coupled into our readout and reduced the total system noise.

The improved sensitivity of the upgraded readout circuit also amplified the low-frequency vibrational backgrounds seen in Run~1, which caused the SQUID amplifier to rail when the magnet was on.  In order to correct this, we implemented an active feedback stabilization (\AFS) system to reduce the low-frequency noise, which is discussed further in the Supplementary Information (SI).

As in Run~1, the magnet and pickup were placed inside a superconducting tin-copper shield and hung from a passive vibration isolation system, consisting of a string pendulum and spring, within an Oxford Instruments Triton 400 dilution refrigerator \cite{ABRA_10cm_Technical}. The magnet and pickup were operated at $\lesssim$1\,K and the SQUIDs were at $\sim$400\,mK, which kept the readout circuit superconducting over the course of the run and kept thermal noise subdominant to SQUID flux noise. Following the procedure of Run~1, the output of the SQUID was run into an 8-bit AlazarTech AT9870 digitizer via a 70\,kHz-5\,MHz bandpass filter. The digitizer was locked to a Stanford Research Systems FS725 Rubidium frequency standard in order to maintain clock accuracy over the coherence time of the axion signal, $\sim1$\,s for signals at 1\,MHz, throughout the data and calibration runs.

\begin{figure}[htb]
    \centering
    \includegraphics[width=\columnwidth]{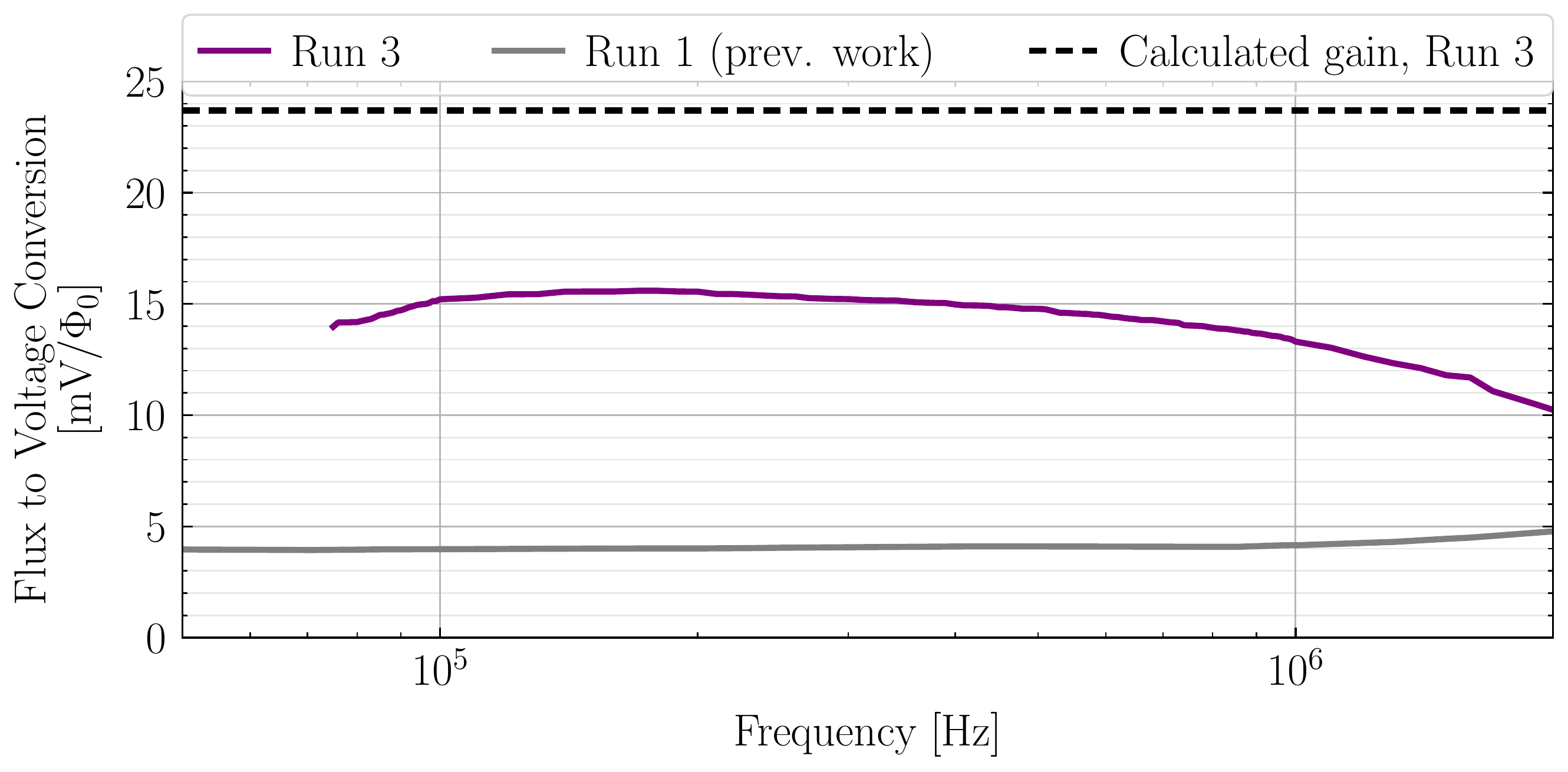}
    \caption{The gain shown here is defined as the change in amplifier output voltage over a corresponding change in input flux amplitude on the pickup cylinder ($\partial V_{\rm out} / \partial \Phi_a$).   Both transfer functions roll off at high frequencies due to the amplifier bandwidth, which we estimate to have a cutoff frequency of approximately 1\,MHz.  We believe the difference in calculated and measured gain is due to inconsistency in the total inductance of the pickup circuit.}
    \label{fig:calib}
\end{figure}

We performed \emph{in situ} magnet-on and magnet-off calibrations in the data-taking configuration by attaching a harmonic signal generator to the calibration circuit and scanning across frequencies and amplitudes. The calibration signal was attenuated and fed into the calibration loop, mimicking the axion effective current signal \Jeff up to geometric factors. The geometry is modeled in COMSOL Multiphysics \cite{COMSOL}, from which we extract the coupling between both the calibration loop and axion effective current signal to the pickup cylinder.  By combining the results of the calibration scans and geometric modeling, we can determine the effective gain, $\partial V_{\rm out}/\partial \Phi_a$, of the SQUID amplifier output voltage as a function of flux on the pickup cylinder (see Fig.~\ref{fig:calib}). This procedure is analogous to that used in Run~1 \cite{ABRA_10cm_Technical}. 

The gain measured by the calibrations for Run~3 differs from the calculated gain by a factor of $\sim$1.8.  By individually calibrating various components of the end-to-end circuit, we determined that this is likely due to a misestimation of the calculated total inductance of the pickup circuit.  The calibrated SQUID noise floors, which set the lower limits of our sensitivity, are shown in Supp. Fig.~\ref{fig:squid_floor}.

\section*{Data collection}

The axion search data was collected using an identical procedure as in Run~1 \cite{ABRA_10cm_Technical}. The SQUID amplifier output voltage was sampled at a frequency of 10\,MS/s, with a $\pm40$\,mV voltage window. The data were stored as a series of power spectral densities (\PSDs), which were computed on-the-fly: \Pten with a Nyquist frequency of 5\,MHz and frequency resolution of $\Delta f=100\,$mHz, \Pone with a Nyquist frequency of 500\,kHz and frequency resolution of $\Delta f=10\,$mHz, and a continuous data stream sampled at 100\,kHz that can be analyzed offline. \Pten (\Pone) is averaged over 800\,s (1600\,s) before being written to disk. In this work, we used the \Pten to search the frequency range from $500\,\mathrm{kHz} - 2\,\mathrm{MHz}$, and the \Pone spectra to search from $50 - 500\,\mathrm{kHz}$.

\section*{Data analysis and results}

An axion signal is expected to manifest as a narrow peak in the \PSD data, as illustrated in Fig.~\ref{fig:detector}.
The width and overall shape of the signal are set by the local \DM velocity distribution, which we take to be the Standard Halo Model with a velocity dispersion of $v_0 = 220$\,km/s and a boost from the halo to the solar rest frame of $v_{\odot} = 232$\,km/s~\cite{Herzog-Arbeitman:2017fte}. With the speed distribution and local DM density fixed, the two free signal parameters are the axion mass, $m_a$, which determines the minimum frequency of the signal, and the coupling $g_{a\gamma\gamma}$, which determines its amplitude through Eq.~\eqref{eqn:coupled_power}. Our analysis procedure closely follows the approach used in the Run~1 search \cite{Ouellet:2018beu,ABRA_10cm_Technical} based on \cite{Foster2018}, which constrains the allowable values of $g_{a\gamma\gamma}$ at each possible value of $m_a$.

The search is performed with a frequentist log-likelihood ratio test statistic (\TS); exact expressions are provided in the SI (see also~\cite{ABRA_10cm_Technical}). 
Our broadband search procedure probes $\sim$ $11.1$ million mass points between $0.41 - 8.27\,\mathrm{neV}$ ($100\,\mathrm{kHz} - 2\,\mathrm{MHz}$) in Run~3.
As we expect only one  axion signal in our search (or at most a small number), the majority of the \TS values are probing the distribution of the null hypothesis. Assuming only Gaussian noise, we expect this null distribution to be a one-sided $\chi^2$-distribution~\cite{Foster2018}, which was indeed the case in Run~1~\cite{Ouellet:2018beu,ABRA_10cm_Technical}. However, the increased sensitivity from the detector upgrades introduced non-Gaussian noise sources that required us to modify our Run~1 analysis procedure. We developed and validated our new procedure on a randomly-selected sample of 10\% of Run~2's $\sim13.7$ million mass points, after which we unblinded the Run~3 data with the procedure fixed.

In Run~1, we searched for an axion signal as a feature appearing above a flat white noise background. For each $m_a$, the search was performed in a narrow window around that mass with the background level allowed to vary independently in each window. For the Run~2 and Run~3 analyses we allow the mean background level of the noise to vary linearly with frequency uniquely in each sliding window.  We use sliding windows of relative width $\delta f / f \approx  5.5 \times 10^{-6}$, starting at $f = (1 -  10^{-6}) \times m_a / (2 \pi)$.

As in Run~1, we use the magnet-off data to veto frequency ranges that also display statistically significant \TS values when $|\mathbf{B}_0| = 0$ and thus the axion power should vanish. However, we observed narrow single-bin `spikes' that appear to drift in frequency on the timescale of our data collection (see Supp. Fig.~\ref{fig:candidates} for an example). 
If interpreted in isolation, these spikes sometimes correspond to statistically-significant excesses. Nevertheless, they are inconsistent with axion signals and are most likely due to unknown environmental noise sources near the detector, persisting throughout Runs 2 and 3; indeed, many of the peaks are distributed at multiples of 50\,Hz. To remove these artifacts, we leverage the fact that the \PSDs are saved periodically to disk yielding a time evolution of the environmental backgrounds; we veto single-bin spikes that move in frequency. We place a 1.0\,Hz veto window around these single-bin spikes. These cuts remove 3.8\% of the axion mass points from our search in the Run~3 data. The magnet-off vetoing procedure removes an additional $0.07$\% of mass points.

After implementing the vetoes, we found the distribution of \TS values in the 10\% Run~2 validation sample deviated from the expected $\chi^2$ distribution; for example, there were 27 mass points with ${\rm TS} > 25$ whereas from the $\chi^2$ distribution we would have expected less than one. 
To account for the deviation in the TS distribution from the $\chi^2$ distribution in a data-driven fashion, we follow the prescription developed and implemented in~\cite{Ackermann:2013uma,Albert:2014hwa,Ackermann:2015lka} for searches for \DM-induced lines in astrophysical gamma-ray data sets. At each mass point, we introduce and profile over a systematic nuisance parameter that would be degenerate with the signal parameter but for a prior that is determined by forcing the \TS distribution to follow the $\chi^2$ distribution. Specifically, we force the \TS distribution to match the null hypothesis distribution at 4$\sigma$ local significance.
This is described further in the SI.

\begin{figure}
    \centering
    \includegraphics[width=\columnwidth]{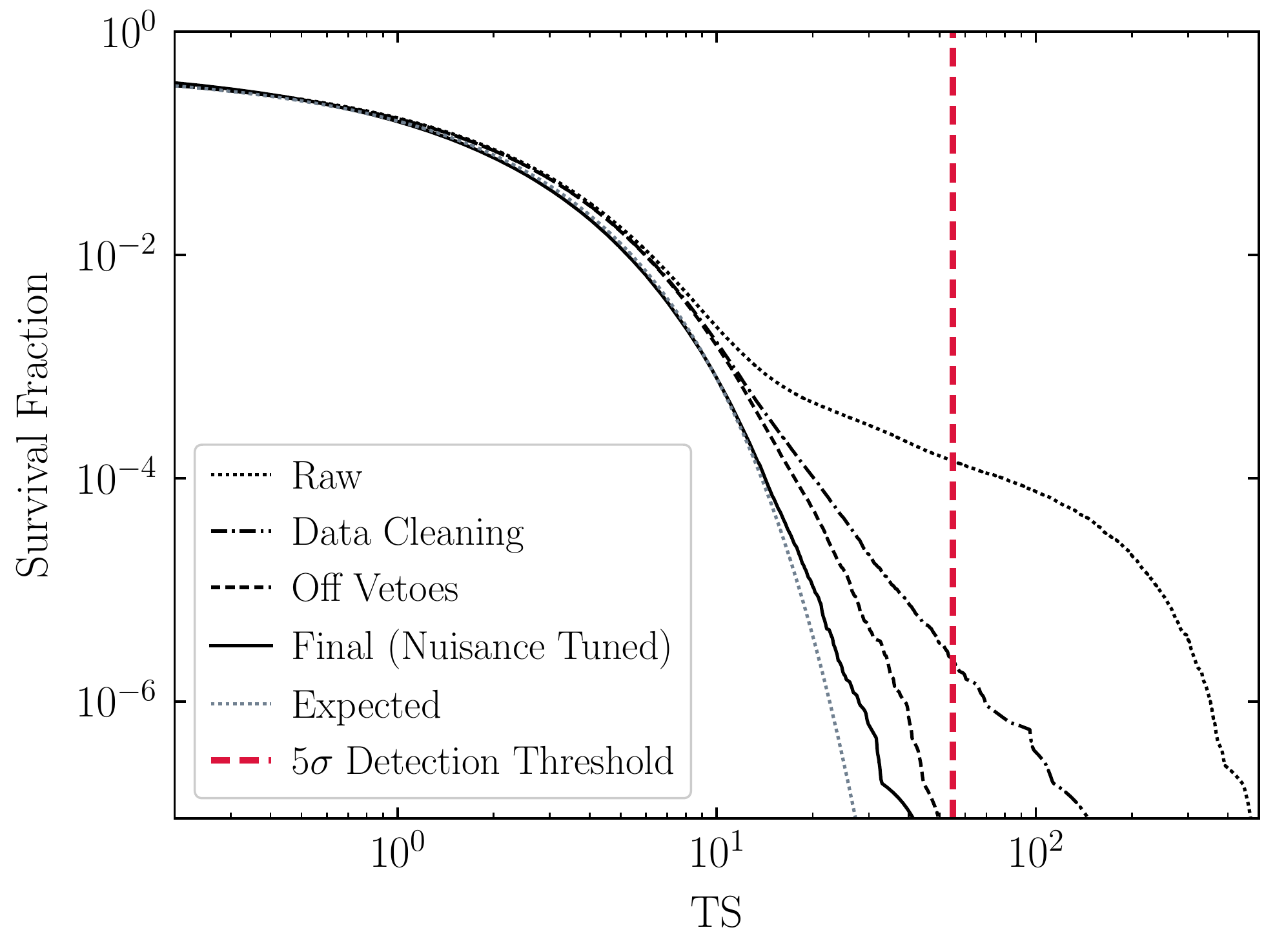}
    \caption{The survival function of TS values from the likelihood analysis of the Run~3 results.  The $y$-axis indicates the fraction of mass points tested with a discovery TS at or above the value on the $x$-axis. Under the null hypothesis, the distribution should follow the survival function of the one-sided $\chi^2$ distribution with one degree of freedom (``Expected,'' dotted gray).  This is indeed the case after data cleaning for {\it e.g.} single-channel excesses in time slices, magnet-off vetoes, and the inclusion of a systematic nuisance parameter, which is tuned in a sliding window at 4$\sigma$ local significance to give the correct number of excesses at or above that significance, masking the signal of interest.  No excesses are found beyond our indicated 5$\sigma$ LEE-corrected discovery threshold.
    }
    \label{fig:TS}
\end{figure}

After the nuisance parameter and vetoing procedures, we construct a likelihood as a function of $g_{a\gamma\gamma}$ at each mass point.
The final distribution of \TS values computed from the likelihoods is shown in Fig.~\ref{fig:TS}; no TS values were found in excess of the 5$\sigma$ look-elsewhere effect-corrected discovery threshold. In the calibration of our analysis procedure, we found one signal candidate in the Run~2 data at over $5\sigma$ global statistical significance (see Supp. Fig.~\ref{fig:candidates}, where a corresponding feature can be seen in the magnet-off data), but that mass point is not significant in the Run~3 analysis.

\begin{figure*}
    \centering
    \includegraphics[width=.95\textwidth]{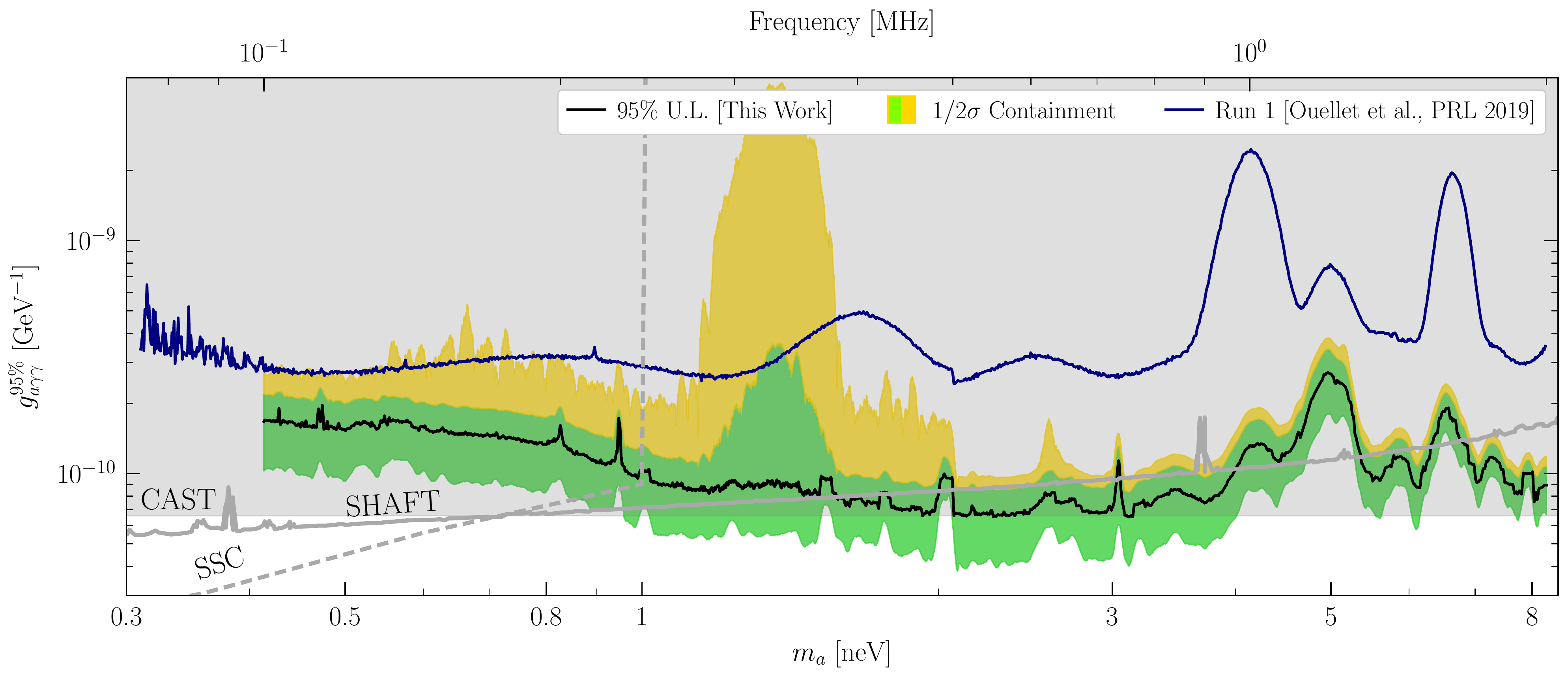}
    \caption{The one-sided 95\% upper limit (U.L.) on $g_{a\gamma\gamma}$ from this work excludes previously unexplored regions of ADM parameter space.  The 1$\sigma$ and 2$\sigma$ containment regions are constructed by taking the appropriate percentiles of the distributions of the limits over narrow mass ranges; note that this means that $\sim$16\% of the upper limits lie at the bottom of the green band. 
    While $\sim$11.1 million mass points are analyzed, the data in the figure are smoothed for clarity.
    {Before smoothing, the strongest limit obtained is $g_{a\gamma\gamma} \lesssim 3.2 \times 10^{-11}~{\rm GeV}^{-1}$ at $m_a \sim 2.99~{\rm neV}$.}
    Our limits surpass those from a number of indicated astrophysical and laboratory searches in this mass range (see text for details).
    }
    \label{fig:limit}
\end{figure*}  

In the absence of an excess consistent with an \ADM origin, we can determine 95\% one-sided upper limits on $g_{a\gamma\gamma}$ as a function of the mass, $m_a$.  The average 95\% upper limits from the Run~3 analysis along with their 1$\sigma$ and 2$\sigma$ expectations under the null hypothesis are indicated in Fig.~\ref{fig:limit}.  In that figure we compare our upper limits to those found from the ADM experiment SHAFT \cite{Gramolin2020a} along with results from the solar axion experiment CAST \cite{CAST2017} and astrophysical $X$-ray searches (SSC)~\cite{Dessert:2020lil}, both of which do not require the axion to comprise the DM. The fraction of vetoed mass points is illustrated in a sliding window in Supp. Fig.~\ref{fig:acceptance}, which also shows the distribution of data fractions included in the analyses. In Supp. Fig.~\ref{fig:RunSystematic} we illustrate the magnitude of the systematic nuisance parameter $g_{a\gamma\gamma}^{\rm nuis.}$, while in Supp. Fig.~\ref{fig:RunLimits_NN} we show what the limit would be without the nuisance parameter tuning. 
Supp. Fig.~\ref{fig:SignalInjection} shows that the 95\% upper limit and discovery TS behave as expected when synthetic axion signals are injected into the real data.

\section*{Discussion}

In this work we present the results from \abra's second physics campaign, searching for \ADM in the mass range 0.41-8.27\,neV.
We find no evidence for \ADM and constrain the axion-photon coupling down to the world-leading level $g_{a\gamma\gamma} \lesssim 3.2\times10^{-11}\,$GeV$^{-1}$ at 95\% confidence.
Our work motivates key elements of the design of future larger-scale experiments. These include the mitigation of stray fields from the magnet and vibrations induced by a modern pulse-tube-based cryogenic system, which limits our current low-frequency reach. The \abra results presented in this Article demonstrate the power of mature simulations for optimizing the design of the detector and for modeling the calibration response. An advanced and novel analysis framework was used to identify noise sources and account for systematic uncertainties in a data-driven fashion. 

Our work identifies three areas that can be addressed in the next physics campaign: (i)  moderate improvements (up to  a factor $\sim$0.4  in $g_{a\gamma\gamma}$)  could  be  achieved  by  further reducing the  wire and SQUID  inductances, (ii) better shielding from environmental noise could increase the sensitivity to $g_{a\gamma\gamma}$ by an order of magnitude at low frequencies, so long as (iii) the fringe fields are reduced or better vibrationally isolated (see Supp. Fig.~\ref{fig:squid_floor}). 
To significantly increase the sensitivity of the experiment, larger magnets with higher fields are needed since the sensitivity to $g_{a\gamma\gamma}$ scales with the detector volume $V$ and field $B_0$ as $\gagg^{-1} \sim B_0V^{5/6}$~\cite{ABRA2016}. The addition of a resonant readout circuit could enhance the reach in \gagg by an additional $\sim$2 orders of magnitude depending on the scanning strategy, with a high frequency readout permitting sensitivity to masses up to 800 neV \cite{ABRA2016,Chaudhuri:2018rqn}. \ABRA is merging with the \mbox{DMRadio} program to realize a series of experiments that chart a path toward discovering the QCD axion in the parameter space corresponding to new physics at the Grand Unification scale~\cite{DOE_BRN,SnowmassOuellet,SnowmassChaudhuri,SnowmassLeder,SnowmassKuenstner}.

\clearpage

\section*{Methods}
The \abra detector is mounted in an Oxford Instruments Triton 400 dilution refrigerator (\DR). The detector is suspended from a vibration isolation system consisting of a $\sim$2\,m long Kevlar thread suspended from a spring. The spring provides isolation in the vertical direction with a rolloff frequency of $f\approx1.4$\,Hz, while the long thread behaves like a pendulum with rolloff frequency of 0.4\,Hz. The detector is thermalized to the coldest stages of the \DR through thin copper strips that cool the detector while minimizing vibration transmission. The fridge was operated without the Turbo Molecular pump in the condensing circuit; all thermometry was disabled and unplugged during data taking to reduce RF noise. 

The data are collected as a series of time averaged \PSDs. The output voltage is digitized at a sampling frequency of 10\,MS/s. It is then Fourier transformed on the fly in 10\,s windows and accumulated into a running average \PSD called \Pten. The data stream is simultaneously down-sampled to a sampling frequency of 1\,MS/s, Fourier transformed in 100\,s windows and accumulated into a running \PSD called \Pone. \Pten (\Pone) is written to disk every 800\,s (1600\,s) and then reset, yielding a detailed time evolution of each spectrum. Two collections, which we refer to as Run~2 and Run~3, were performed. Run~2 contained 1,168,000\,s (324\,h) of magnet on data in 1,460 \Pten spectra and 700 \Pone spectra and 320,000\,s (89\,h) of magnet off data in 388 \Pten spectra and 196 \Pone spectra. Run~3 contained 1,091,200\,s (303\,h) of magnet on data in 1,364 \Pone spectra and 682 \Pone spectra and 448,000\,s (124\,h) of magnet off data in 560 \Pten spectra and 280 \Pone spectra. Due to differences in the readout configuration, Run~2 was not as sensitive as Run~3 and the results are not presented here, though the Run~2 data are used for tuning the analysis procedure.

The data were analyzed in search of the expected axion line-shape using a profiled Gaussian likelihood with a linear background model, estimating the data variance as a floating nuisance parameter at 11.1 million axion mass points between 100 kHz and 2 MHz. Data from Run~3 was analyzed independently based on an analysis validated on 10\% of Run~2 data. The data were cleaned by removing spectral excesses confined to a single frequency bin. The analysis was initially applied to independently divided time-continuous subintervals of the data (each of size approximately 5\% of the full data volume), allowing for a filtering of transient excesses applied independently at each mass. Data which passed the filtering were then stacked and analyzed for each run,
then joined to produce a 95$^\mathrm{th}$ percentile upper limit and detection significance. The detection significance was then corrected by a Gaussian penalty term with a hyperparameter to accommodate systematic effects that may produce spurious detections.

\section*{Data availability}
Source data for this paper are made publicly available;\footnote{\url{https://github.com/joshwfoster/ABRA_Results_2020}} all other data may be made available upon reasonable request.

\section*{Code availability}
The code that supports the results presented in this paper may be made available by the corresponding authors upon reasonable request.

\begin{acknowledgments}
{\it 
We would like to thank Kent Irwin and our \mbox{DMRadio} colleagues for useful discussions and look forward to the next-generation experiment. We would like to thank those that took part in Run 1 of ABRACADABRA-10cm including  Zachary Bogorad, Janet Conrad, Joseph Formaggio, Joe Minervini, Alexey Radovinsky, Jesse Thaler, and Daniel Winklehner. We thank Christopher Dessert for useful analysis discussion. This  research  was  supported  by  the  National  Science  Foundation  under  grant  numbers  NSF-PHY-1658693,  NSF-PHY-1806440. J.F. and B.R.S.  were  supported  in  part  by  the  DOE  Early Career  Grant  DESC0019225, through  computational resources  and  services  provided  by  Advanced  Research Computing  at  the  University  of  Michigan,  Ann  Arbor, and by computational resources at the Lawrencium computational cluster provided by the IT Division at the Lawrence Berkeley National Laboratory, supported by the Director, Office of Science, and Office of Basic Energy Sciences, of the U.S. Department of Energy under Contract No.  DE-AC02-05CH11231. Y.K. is supported in part by US Department of Energy grant DE-SC0015655. R.N. is supported by the National Science Foundation Graduate Fellowship under Grant No. DGE–1746047. N.L.R. is supported by the Miller Institute for Basic Research in Science at the University of California, Berkeley. C.P.S. is supported in part by the National Science Foundation Graduate Research Fellowship under Grant No. 1122374. R.H. and K.R. are supported by the U.S. Department of Energy, Office of Science, Office of Nuclear Physics under Awards No. DEFG02-97ER41041 and No. DEFG02-97ER41033. We would like to thank the University of North
Carolina at Chapel Hill and the Research Computing group
for providing computational resources and support that
have contributed to these research results.
}

\end{acknowledgments}

\clearpage

\onecolumngrid
\begin{center}
  \textbf{\large Supplementary Material for the search for low-mass axion dark matter with ABRACADABRA-10\,cm}\\[.2cm]
  \vspace{0.05in}
  {Chiara P. Salemi, \ Joshua W. Foster, \ Jonathan L. Ouellet, \ Andrew Gavin, \ Kaliro\"e~M.~W.~Pappas, Sabrina~Cheng, \ Kate A. Richardson, \ Reyco Henning, \ Yonatan Kahn, \ Rachel Nguyen, \ Nicholas L. Rodd, \ Benjamin R. Safdi, \ and \ Lindley Winslow}
\end{center}

\twocolumngrid
\setcounter{equation}{0}
\setcounter{figure}{0}
\setcounter{table}{0}
\setcounter{section}{0}
\setcounter{page}{1}
\makeatletter
\renewcommand{\theequation}{S\arabic{equation}}
\renewcommand{\thefigure}{S\arabic{figure}}
\renewcommand{\thetable}{S\arabic{table}}

\section{Detector upgrade and electromagnetic simulations}

The sensitivity of \abra to \ADM is set by the coupling strength between the axion induced current \Jeff and the readout SQUIDs. This coupling can be conceptually split into two parts: the coupling between \Jeff and the pickup, and the coupling between the pickup and the SQUIDs. Before Run~2, the \abra detector was upgraded in two ways to increase each of these two coupling strengths.

The first step of the upgrade was the installation of the superconducting pickup cylinder. The pickup cylinder geometry more effectively cancels the flux induced by \Jeff and thus couples more strongly to it. The cylinder was constructed out of a 150\,$\mu$m thick sheet of Nb wrapped around a PTFE tube, secured with Kapton tape.  The resulting cylindrical pickup was 10\,cm tall with a 5.1\,cm diameter and centered vertically in the magnet bore. This is close to the maximum diameter that could practically fit. A 1\,mm gap was left in the wrapping of the Nb sheet to prevent electrical contact and the formation of a complete loop. The PTFE tube was glued and clamped onto the magnet support structure inside the superconducting shielding can. From experience in Run~1, a strong mounting was critical to reducing relative motion between the pickup and magnet.

The second step of the upgrade was a replacement of the wiring between the pickup cylinder and the SQUID readouts. The new wiring -- along with the new pickup cylinder -- reduced the total inductance of the readout circuit, resulting in more current in the SQUIDs. The new wiring consists of 75\,$\mu$m  superconducting solid NbTi twisted-pair wires that are spot welded to two corners of the Nb sheet. Spot welding ensures a superconducting connection between the Nb sheet and the NbTi wires. We used a series of four spot welds on each corner for redundancy, in case of breakage during handling or due to differential thermal contraction. These wires were then taped to the PTFE cylinder with Kapton in order to minimize stress on the connections. The wires run $\sim$57.5\,cm to the SQUID input. The wires are shielded in a superconducting capillary which extends from about 1\,cm from the Nb sheet up to the point where the wires enter the SQUID shielding can. In addition to providing electric shielding, this capillary also reduced the inductance per unit length. The inductance of the new cylinder and the readout wires are calculated using simulations in COMSOL Multiphysics to be 20\,nH and 288\,nH, respectively. This decrease in the readout inductance was confirmed with calibrations up to a factor $\sim$1.8.

\section{Active Feedback System}

A major challenge encountered in Run~1 was low frequency vibrations converting stray fields from the magnet into low frequency noise. Since this noise is generally below the frequency ranges of interest, we were able to simply filter it and ignore it. In Runs 2 and 3, the higher gain of the upgraded detector amplified this vibrational noise enough to rail the SQUID amplifier. Because most of this noise was relatively slow -- below $\sim1$\,kHz -- we installed an active feedback system to cancel it. We fed part of the output signal into the input of a Stanford Research Systems SIM960 analog PID controller. The output was filtered through a 1\,kHz low-pass filter (LPF) and fed into the calibration loop via a 10\,dB warm attenuator followed by 40\,dB of cold attenuation, as sketched in Fig.\,\ref{fig:circuits}. The LPF guaranteed that the feedback system could not interfere with signals in our ROI, while the warm 10\,dB attenuator reduced the power dissipated on the cold stages of the fridge. 

Before Run~3, we added power combiners and power splitters to the active feedback circuit in order to better impedance match and isolate the various parts of the circuit.  This improved our in situ calibration, as described below.

\section{Detector calibration}

\begin{figure}
    \centering
    \includegraphics[width=.45\textwidth]{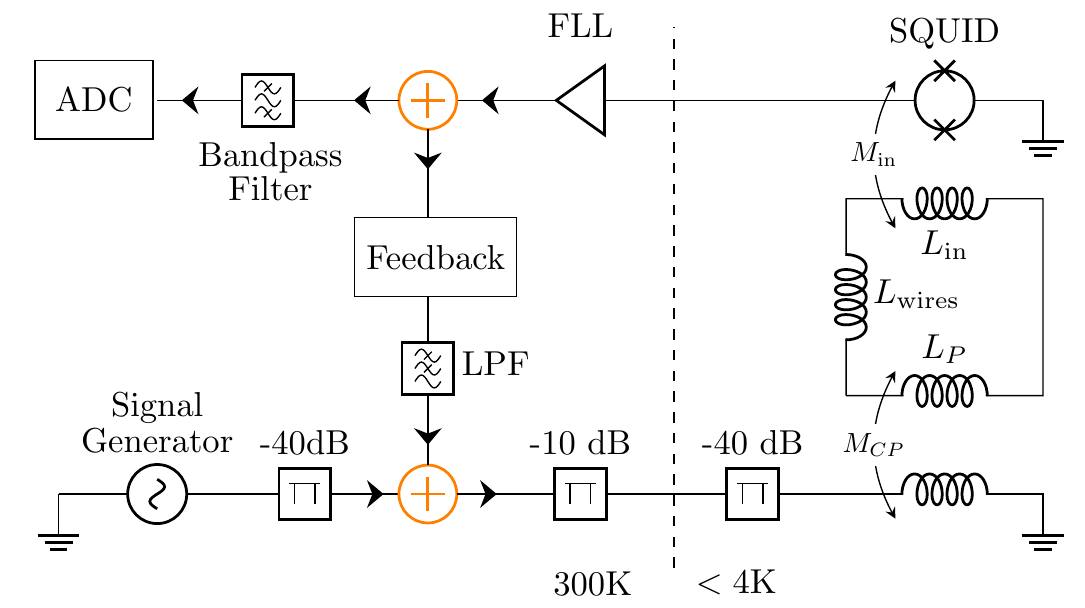}
	\caption{\abra Run~3 calibration circuit diagram. A fake axion signal generated in the signal generator is attenuated by 93\,dB (including 3dB loss in the combiner) before being coupled into the pickup cylinder analogously to an axion signal. The resulting power excess is readout on the SQUID and measured in the ADC digitizer. In Run~3, calibration is performed with the magnet turned on and the active feedback circuit running. During data taking, the signal generator is replaced with a 50\,$\Omega$ terminator. The flux-lock feedback loop (FLL) feedback resistor and inductor are omitted for clarity.}
    \label{fig:circuits}
\end{figure}

At a basic level, the \abra readout converts the flux through the pickup cylinder $\Phi_p$ to an output voltage from the SQUID amplifier, $V_{\rm SQUID}$. The detector calibration provides an end-to-end measurement of the detector response $\partial V_{\rm SQUID}/\partial \Phi_p$ to an axion-like signal across the full range of frequencies being searched. A schematic of the detector configuration for the Run~3 calibration can be seen in Fig~\,\ref{fig:circuits}. We generate a fixed frequency signal of known amplitude using an Stanford Research Systems SG380 signal generator. This signal is attenuated by 93 dB before passing into the calibration loop in the detector. The current in the calibration loop generates a flux through the pickup cylinder, inducing a current and response in the readout circuit in the same way that an axion signal would, up to geometric factors. 

The response of the system to a calibration signal can be written as 
\begin{equation}
\frac{\partial V_{\rm ADC}}{\partial V_{\rm Sig}} = \frac{\partial V_{\rm ADC}}{\partial V_{\rm SQUID}}\frac{\partial V_{\rm SQUID}}{\partial \Phi_p}\frac{\partial \Phi_p}{\partial I_C}\frac{\partial I_C}{\partial V_{\rm Sig}}
\end{equation}
where $V_{\rm ADC}$ is the RMS voltage measured by the digitizer, $I_C$ is the
RMS current entering the calibration loop, and $V_{\rm Sig}$ is the peak-to-peak voltage output by the signal generator. The first and last terms in this conversion are determined by the warm electronics and cold attenuators, and can be measured directly, while the third term is the mutual inductance between the calibration loop and pickup cylinder, which is modeled in COMSOL. By dividing the measured end-to-end calibration by these three terms, we are left with the resulting flux to voltage conversion of the \abra readout circuit $\partial V_{\rm SQUID}/\partial \Phi_p$. 

During Run~3, the calibration was performed in an identical configuration to data taking, namely with the magnet on and the \AFS active. The resulting calibration can be seen in Fig.~\ref{fig:calib}, and agreed very well with our calculated signal gain. The rolloff above $\sim$1\,MHz corresponds to the finite bandwidth of the SQUID electronics. The Run~3 calibration circuit diagram is provided in Fig.~\ref{fig:circuits}.  During Run~2, we were not yet able to accurately calibrate the detector with the \AFS system in place, which led to the decision to not present this data in our limits.
Instead, the uncalibrated Run~2 data was used to tune our analysis procedure.

\begin{figure}
    \centering
    \includegraphics[width=.49\textwidth]{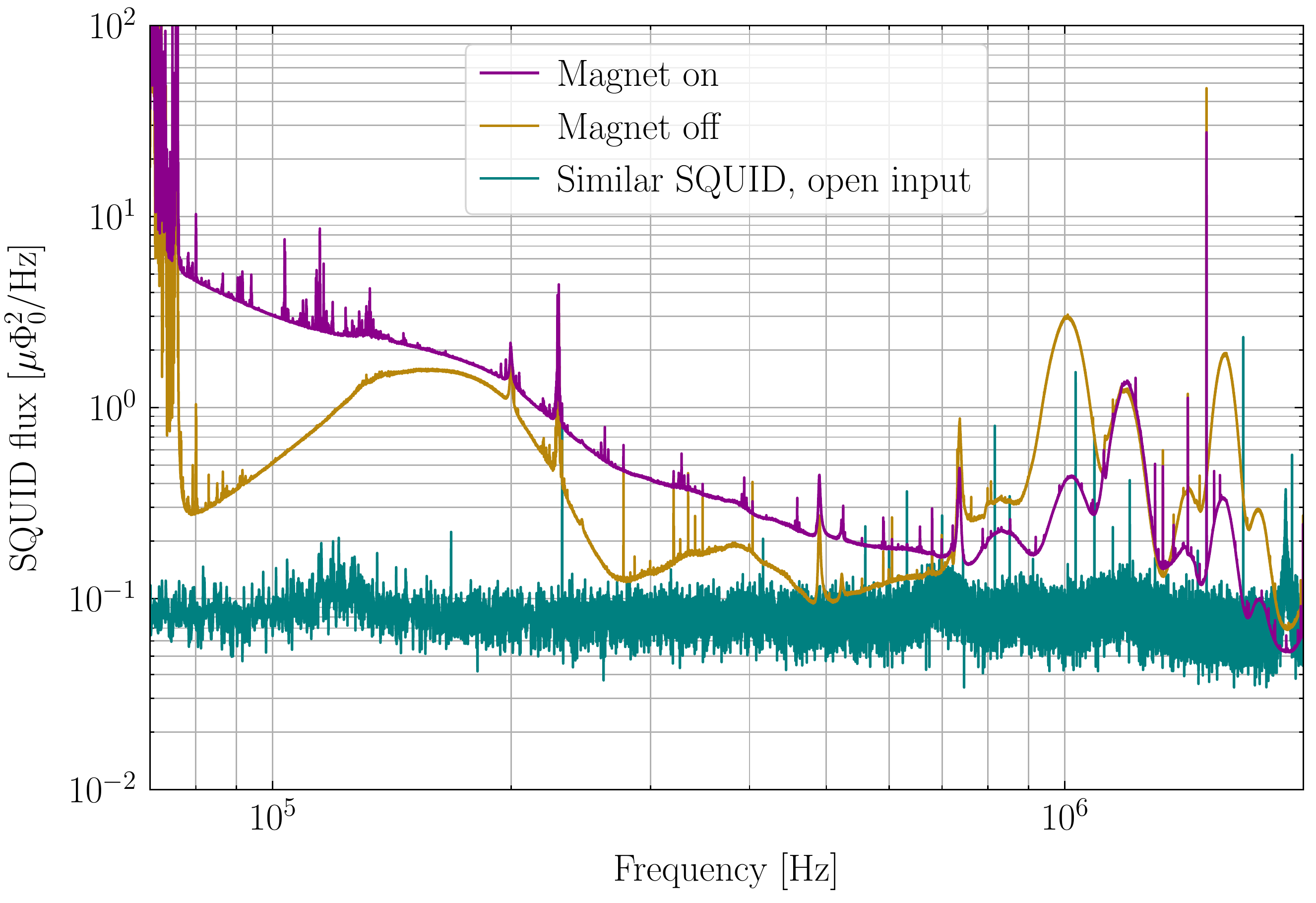}
    \caption{The SQUID flux for Run~3 over the 70\,kHz to 2\,MHz frequency range at which we collect data. The magnet on noise level (magenta) is elevated compared to data taken with the magnet off (gold) primarily due to vibrating fringe magnetic fields.  For comparison, the noise level from a similar SQUID without anything plugged into its input is plotted in teal.}
    \label{fig:squid_floor}
\end{figure}

The flux noise determines the lower limit of our sensitivity.  In particular, the 95\% upper limit on $g_{a\gamma\gamma}$ under the null hypothesis scales like the square root of the flux noise, which is shown in
Fig.~\ref{fig:squid_floor}.  In that figure we illustrate three different noise levels through the SQUID: (i) the measured magnet on flux, which is the relevant flux for the axion signal analysis; (ii) the magnet off flux; (iii) the flux measured in a similar SQUID that is not connected to the pickup loop circuit (labeled ``open input").  The increased noise level in the magnet off data relative to the open input SQUID is likely the result of imperfect shielding, with environmental noise magnified by the pickup loop.  On the other hand, when the magnet is on increased noise is apparent at low frequencies, which is the result of the magnetic fringe fields giving frequency-dependent flux noise because of vibrations.  Increasing the quality of the shielding and decreasing either the magnitude of the fringe fields or their vibrational coupling to the pickup loop would improve the sensitivity.  

\section{Likelihood analysis}

In this section, we describe the implementation of the analysis framework used to produce upper limits on $\gagg$ and determine detection significances for potential excesses. We first define the profiled Gaussian likelihood used herein, followed by our procedure for cleaning the data to enable the removal of spurious excesses and confounding backgrounds. We then detail our treatment of a nuisance hyperparameter used to address potential systematics in the data, describe our results in terms of survival functions and upper limits on $\gagg$, and demonstrate the efficacy of our analysis pipeline with injected signal tests.  Though uncalibrated and not presented here, the Run~2 data was used while constructing the likelihood analysis framework. As such, we include it in the discussion below.

\subsection{Likelihood for Axion Signal Detection}

The likelihood analysis utilized in this work is performed using the signal modeling formalism developed in \cite{Foster2018}, which was also used in studying the Run~1 results~\cite{Ouellet:2018beu, ABRA_10cm_Technical}.
Our starting point is a series of $N$ samples of the flux in the pickup loop $\{\Phi_n\}$, made over a collection time $T$ and at a sampling frequency $f=1/\Delta t$ (such that $N \Delta t = T$). 
In the presence of an axion, this flux will receive a contribution from both the \DM signal and any background.  The mean expectation for the PSD at a frequency $f_k = k/T$ is
\begin{equation}
\Phi_k = A(\gagg) s_k(m_a) + \mu_k\,,
\end{equation}
with $\mu_k$ is the mean expected background at this frequency.  The signal strength parameter $A$ is given in \eqref{eqn:coupled_power} and is controlled by the unknown $\gagg$, while $s_k$ is the signal template for a specific axion mass:
\begin{eqnarray}
s_k(m_a) = 
\left\{ \begin{array}{ll}
\frac{\pi f(v_{\omega})}{m_a v_{\omega}} & f_k > m_a/2\pi\,, \\
0 & f_k \leq m_a/2\pi\,.
\end{array} \right.
\label{eq:skdef}
\end{eqnarray}
Here, $v_{\omega} = \sqrt{4\pi f_k/m_a - 2}$ and $f(v)$ is the local axion speed distribution, which we take to be the Standard Halo Model with boost velocity $v_{\odot} = 232$ km/s and velocity dispersion $v_0 = 220$ km/s.
To the extent the background is Gaussian in the time domain, the PSD formed from this data will be exponentially distributed, and the sum of multiple PSDs formed during data stacking will be Erlang-distributed.
Nevertheless, in the limit of a large number of stackings, the Erlang-distribution becomes normally-distributed.
For this reason we are justified in analyzing the data using a Gaussian likelihood.

In detail, the likelihood used is given by
\es{eq:likelihood}{
&\mathcal{L}(d | m_a, A;\, \mathbf{a}, \sigma) \\
= &\prod_k \frac{1}{\sqrt{2 \pi \sigma^2}}\exp \left[ - \frac{(d_k - A s_k - \mu_k(\mathbf{a}))^2}{2 \sigma^2} \right],
}
where $d_k$ is the average stacked data, $A$ and $s_k$ determine the axion signal as described above, $\mu_k$ is the background model (specified by parameters $\mathbf{a}$), and $\sigma$ is the standard deviation which we will treat as a nuisance parameter, and therefore estimate directly from the data.
For a given axion mass $m_a$, the signal only has support over a narrow frequency range, and therefore we truncate the likelihood to $k$ values between $m_a (1-(v_{\odot} + v_0)^2 / 2)/2\pi$ and $m_a (1+2(v_{\odot} + v_0)^2)/2\pi$.
Over this narrow range, we find the background is adequately described by a first order polynomial, defined by the two-component vector $\mathbf{a}$ (c.f. Run~1 where the background in each signal window was described by a flat white-noise spectrum).
In summary, our likelihood is a function of five parameters: $m_a$ and $A$, which define the location and normalization of the signal, and nuisance parameters $\mathbf{a}$ and $\sigma$, which describe the mean size, slope, and fluctuations of the background.

Our goal is to use the likelihood in \eqref{eq:likelihood} to search for deviations from the background only distribution indicative of the presence of an axion.
To do so we define the following test statistic (TS), which is a log-likelihood ratio of the signal and null models,
\begin{equation}
t(m_a, A) = 2 \ln \left[\frac{\mathcal{L}(d | m_a, A;\, \hat{\mathbf{a}}, \hat{\sigma})}{\mathcal{L}(d | m_a, A=0;\, \hat{\mathbf{a}}, \hat{\sigma})} \right].
\label{eq:TS}
\end{equation}
Hatted background quantities are fixed to the value at which the likelihood attains its maximum value, given the specified signal values (i.e. for $A \neq 0$, $\hat{\mathbf{a}}$ and $\hat{\sigma}$ will in general take different values in the numerator and denominator).
In other words, in defining this TS, we profile over the background nuisance parameters.
The above test statistic is defined for any $m_a$ and $A$.
For a given $m_a$, we then define the discovery \TS as 
\begin{equation}
\mathrm{TS}(m_a) = \max_A t(m_a, A).
\label{eq:maxTS}
\end{equation}
The maximization of $A$ is initially performed over a range including positive and negative values, which is critical for the valid interpretation of $\mathrm{TS}$ as a $\chi^2$-distributed quantity under Wilks' theorem; intuitively, background fluctuations below the mean are just as likely as those above.
However, as the presence of an actual axion signal will only result in positive spectral excesses, we take $\mathrm{TS}(m_a) = 0$ when the test statistic is maximized with $A < 0$. 
Accordingly, the discovery test statistic is expected to have the following asymptotic distribution
\begin{equation}
p(\mathrm{TS}) = \frac{1}{2} \left[ \delta(\mathrm{TS}) + \chi^2_{k=1}(\mathrm{TS})\right],
\label{eq:halfChi}
\end{equation}
which is expressed in terms of $\chi^2_{k=1}$, the probability density function for the $\chi^2$-distribution with one degree of freedom, and a Dirac $\delta$ function.

Using the test statistic in \eqref{eq:TS}, we search for evidence of \ADM with masses $m_a$ such that the signal would appear within the frequency range $f_{\rm min}=100$ kHz to $f_{\rm max}=2$ MHz.
The local significance of any excess can be quantified by inverting the distribution in \eqref{eq:halfChi}.
In order to cover our entire frequency between $f_{\rm min}$ and $f_{\rm max}$, this search is performed in $11.1$ million signal windows.
As such, the local significance can be misleading and we should instead interpret results after accounting for the look-elsewhere effect (LEE).
In doing so, we also need to account for the fact that due to the finite extent of the axion signal template $s_k$, nearby windows are correlated.
We account for this self-consistently using the formalism developed in~\cite{Foster2018}, from which we compute the $N\sigma$ detection threshold, accounting for the LEE by
\begin{equation}
\mathrm{TS}_\mathrm{thresh}(N) = \left[\Phi^{-1}\left(1- \frac{4 v_0^2\, \Phi(N)}{3 \ln (f_{\mathrm{max}}/f_\mathrm{min})}\right) \right]^2,
\end{equation}
expressed in terms of $\Phi$, the cumulative density function of the zero mean and unit standard deviation normal distribution, with $\Phi^{-1}$ its inverse.
Using this formalism, we find that the $5\sigma$ detection threshold accounting for the LEE is $\mathrm{TS}_\mathrm{thresh} \approx 55$.

A direct application of the formalism outlined thus far to the Run~2 and 3 data sets would result in a number of excesses at moderate or even high significance.
Rather than interpreting this result as the discovery as a large number of \ADM signatures, we interpret these as false positives sourced by coherent backgrounds that are not adequately captured by our null model.
We employ two strategies for improving the background model in light of this.
Firstly, we apply a data cleaning procedure in order to remove excesses inconsistent with that expected for \ADM, for example transient spectral features or features that appear also when the magnet is off.
Secondly, after applying our data cleaning pipeline, we modify our likelihood with a nuisance parameter tuned against the ensemble of observed significance values in the clean dataset; a data driven method for ensuring the quoted significance is consistent with the distributions observed directly in data.
After applying both corrections factors, we find no significant excesses remain in our combined dataset. 
We now describe each of these strategies for improving our background treatment in more detail.
We emphasize that all data cleaning and analysis procedures
were developed and tested on the 10\% of the Run~2 data which was unblinded and was then applied identically to the Run~3 data.
The 10\% was obtained by subdividing the full Run~2 dataset into 1,000 frequency subsets of equal size, and then taking the first 10\% of each subset in frequency.

\subsection{Data Cleaning Procedure}

Many of the excesses present in the uncleaned data are characterized by a narrow spectral feature, often present in a single frequency bin.
The features often drift throughout our collection time, and appear at regular frequency intervals.
Although such features are inconsistent with the axion signal expectation, which should be distributed over several frequency bins, such narrow features are far more consistent with our signal model than our linear background model, and therefore result in high-significance $\mathrm{TS}$ values.

A notable example is the background resulting from AM radio broadcasts: these manifest as large excesses at uniform 10 kHz intervals from 560 kHz to 1.60 MHz.
We identify these AM radio signals in our data and remove them with a mask of width 15 Hz centered on the radio signal peak, beyond which the radio signal falls below our noise floor.
In other cases the origin of the features is unclear, although the fact that many appear at 50 Hz intervals suggests a universal environmental origin.
Regardless, we remain agnostic to their origin and instead remove them using a data-driven procedure we now outline.

For each frequency bin, we identify a single-bin excess as follows.
We determine the mean and standard deviation of the data on either side of the bin of interest; in particular, we use the 10 bins on both sides, ignoring the immediately adjacent frequencies.
We then use these results to calculate the significance of the data observed in the bin of interest.
We repeat this procedure for each frequency bin in each independently collected dataset, i.e. the data before it is stacked.
If the bin of interest attains a significance $\delta \chi^2 > 100$ in any one dataset, then it is flagged for masking.
If the bin is not flagged by this procedure, we then stack the dataset and repeat this procedure once more.
If after stacking, the bin now has $\delta \chi^2 > 35$, then it is again flagged.
For all flagged bins, we mask the 21 frequencies centered on the bin of interest.
The motivation for considering the individual and stacked datasets is to identify both excesses that drift with time and also those that are only significant in the stacked data where we perform our fiducial analysis.

\begin{figure*}
    \centering
    \includegraphics[width=.45\textwidth]{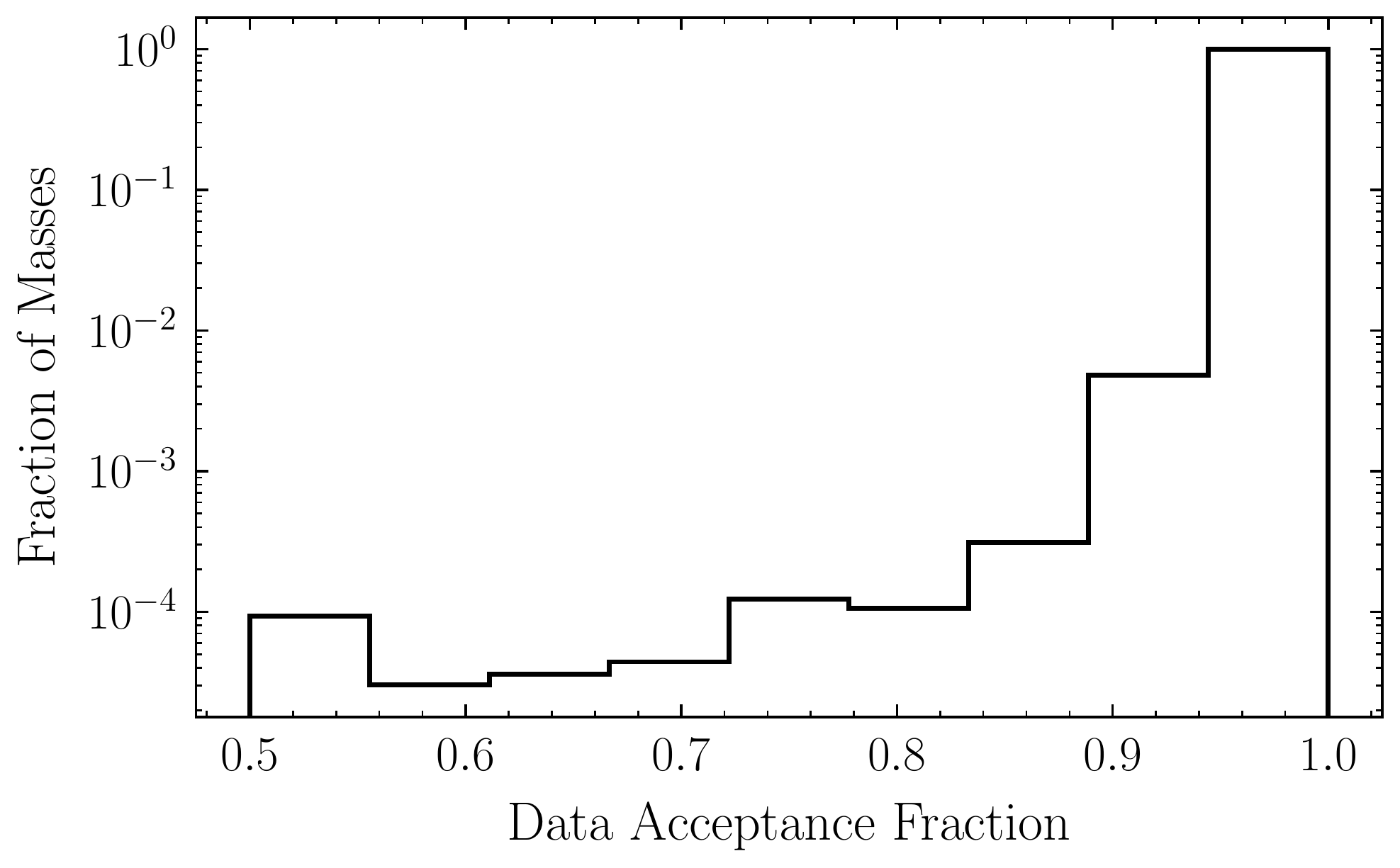}\hspace{3ex}\includegraphics[width=.45\textwidth]{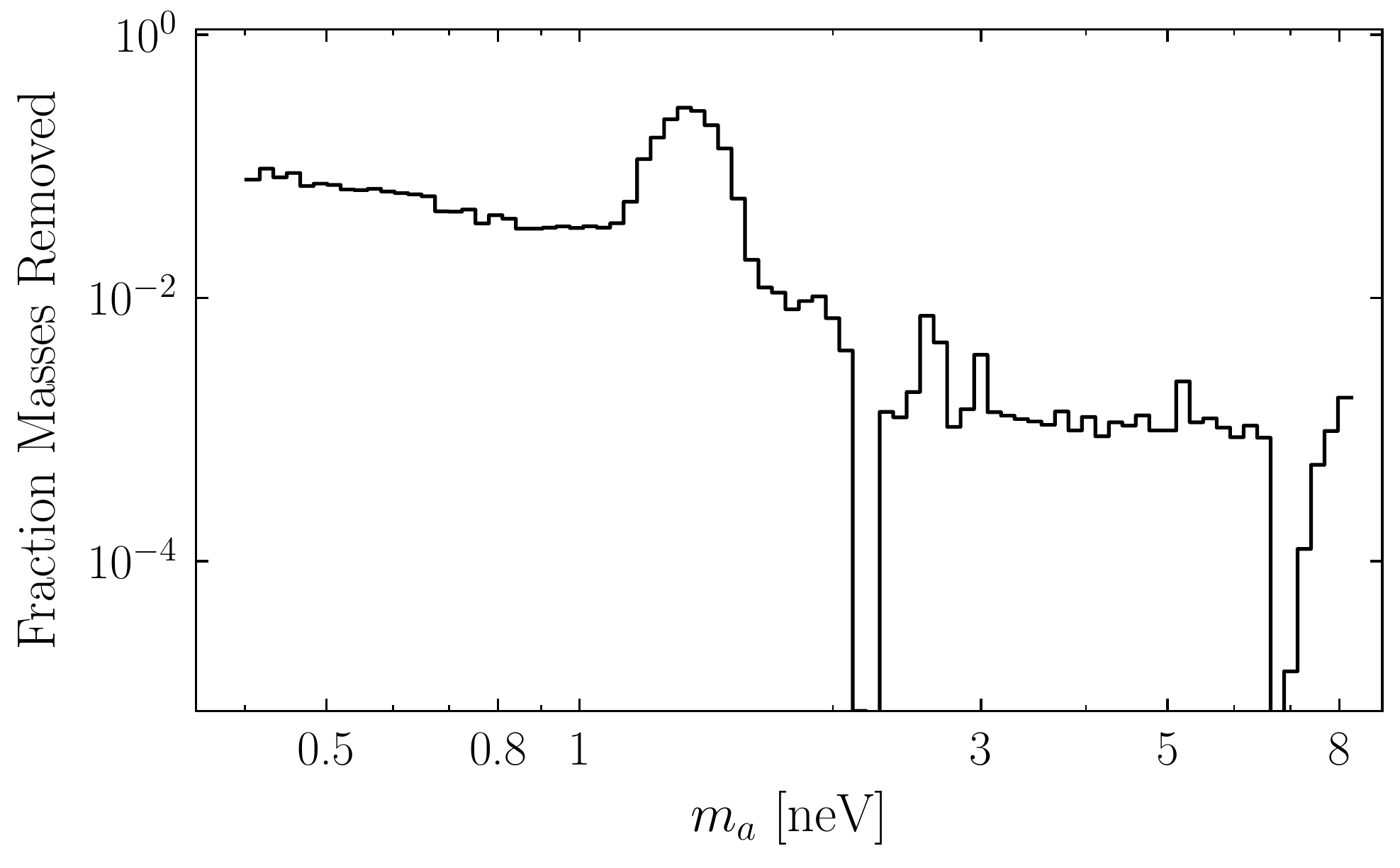}
    \caption{ (\textit{Left}) The histogrammed data acceptance fraction under the data filtering over all masses analyzed in Run~3 data. (\textit{Right}) The fraction of masses removed by magnet off vetoes as a function of frequency in Run~3 data. The acceptance fraction is determined within 100 log-spaced bins between the minimum and maximum axion masses within our analysis range.  Note that while we display the Run~2 results, those were used only to develop our analysis protocols and not in the physics analysis.}
    \label{fig:acceptance}
\end{figure*}

After the single-bin spikes have been identified and removed, we perform an initial analysis of the data. We analyze the \Pten (\Pone) data providing a frequency resolution of $0.1$ ($0.01$) Hz for axions which would produce a signal in the 500 kHz - 2 MHz (50 - 500 kHz) frequency range. We stack the  Run~2 \Pten (\Pone) data, which consists of 1460 (700) spectra, into 20 subintervals, each of which are initially analyzed independently. For each axion mass, each subinterval is analyzed independently, with the 50\% of subintervals which realize the smallest values of the TS for discovery and any additional subintervals which have $\mathrm{TS} < 9$ accepted. The accepted subintervals are then stacked into a single dataset and analyzed. An analogous procedure is applied to the Run~3 data, where the \Pten (\Pone) data, consisting of 1364 (682) spectra, are divided into 22 subintervals. This TS filtering procedure was implemented in order to mitigate the impact of transient excesses that imitate an axion signal in some of the subintervals and might produce a spurious excess if included in the stacked data. With this exclusion criteria, under the null, each spectrum is expected to be excluded with probability $0.1\%$, making this a relatively conservative exclusion criterion, although it does have the effect of somewhat weakening the detection sensitivity of the analysis. The binned data acceptances for the complete Run~3 analysis are shown in the left panel of Fig.~\ref{fig:acceptance}, which show that all data is accepted into the stacked analysis data for the overwhelming majority of mass points.

\begin{figure*}
    \centering
    \includegraphics[width=.55\textwidth]{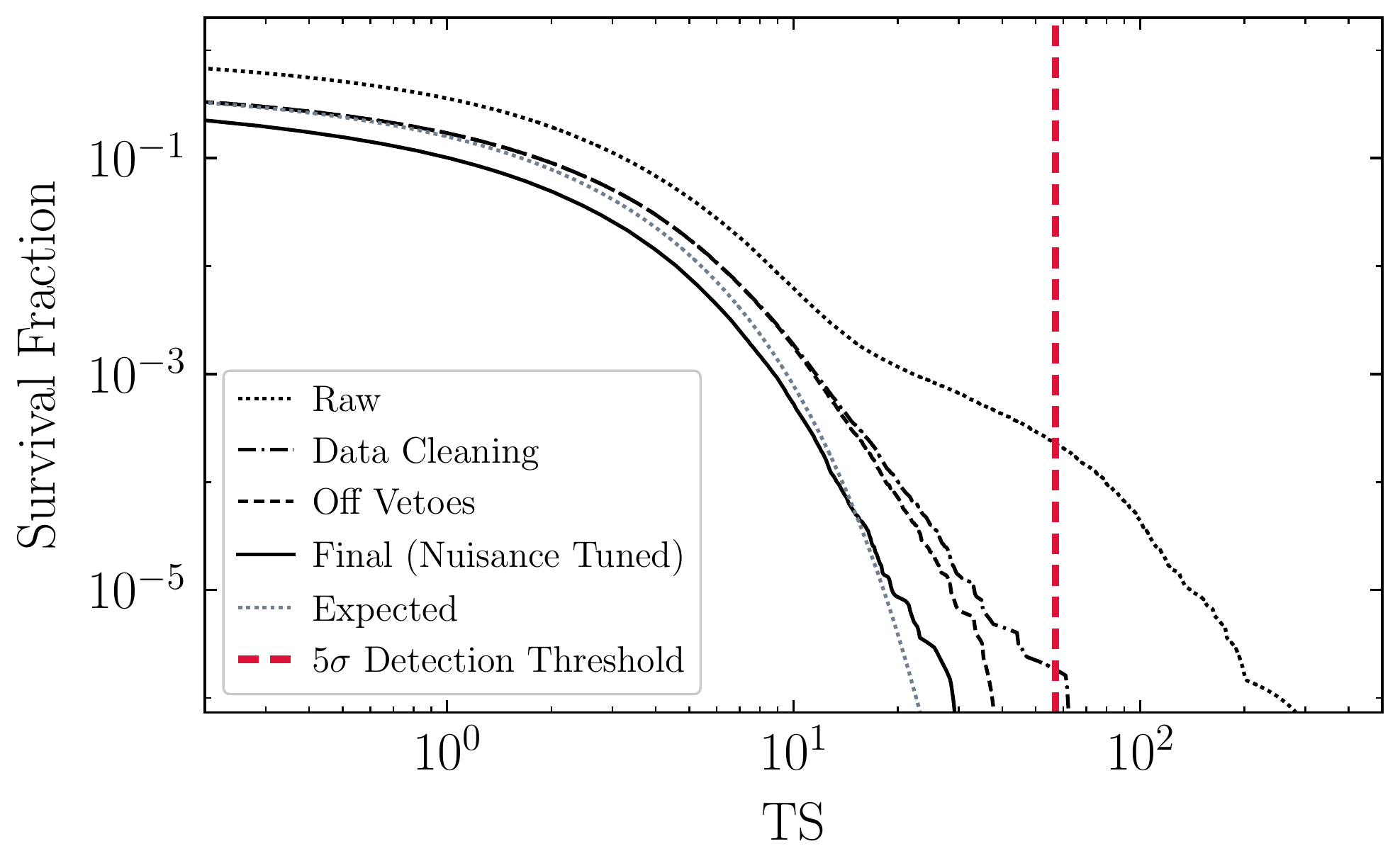}
    \caption{As in Fig.~\ref{fig:TS}, but evaluated on the 10\% of unblinded Run~2 data against which we calibrated our analysis procedure.}
    \label{fig:TestingSurvivals}
\end{figure*}  

Finally, we perform a series of vetoes aimed at removing any remaining excess which have their origins in unmodeled backgrounds or instrumental effects. We directly analyze stacked, unfiltered \Pten and \Pone data which are collected with the magnet off. Since no axion detection can be made with the magnet off, any mass points which are excesses at $\mathrm{TS} > 16$ in both the magnet-on and magnet-off data are vetoed. 

\subsection{Nuisance Parameter Correction}

After applying both our individual bin flagging and vetoing procedures, the remaining dataset is designated clean.
Nevertheless, the distribution of \TS values remains inconsistent with that expected for the asymptotic one-sided $\chi^2$ distribution given in \eqref{eq:halfChi}, indicative of further background mismodeling.
To resolve this, we implement an additional nuisance parameter correction to our likelihood.

In detail, we modify the likelihood and \TS with additional nuisance parameters $A_m$ and $\sigma_{A_m}$ as follows,
\begin{widetext}
\begin{equation}
\mathcal{L}(d |m_a, A;\,\mathbf{a}, \sigma, A_m,\sigma_{A_m}) = \mathcal{N}(A_m | 0, \sigma_{A_m}) \prod_k \frac{1}{\sqrt{2 \pi \sigma^2}} \exp \left[ - \frac{(d_k - (A+A_m) s_k - \mu_k(\mathbf{a}))^2}{2 \sigma^2} \right]
\label{eq:NuisanceLikelihood}
\end{equation}
\begin{equation}
\mathrm{TS}(m_a | \sigma_{A_n}) 
= 2 \ln \left[ \frac{\max_A \mathcal{L}(d |m_a, A;\,\hat{\mathbf{a}}, \hat{\sigma}, \hat{A}_m,\sigma_{A_m})}{\mathcal{L}(d |m_a, A=0;\,\hat{\mathbf{a}}, \hat{\sigma}, \hat{A}_m,\sigma_{A_m})} \right].
\label{eq:NuisanceTS}
\end{equation}
\end{widetext}
The index $m$ indicates that the nuisance parameters depend on the signal window under consideration.

\begin{figure*}
    \centering
    \includegraphics[width=.85\textwidth]{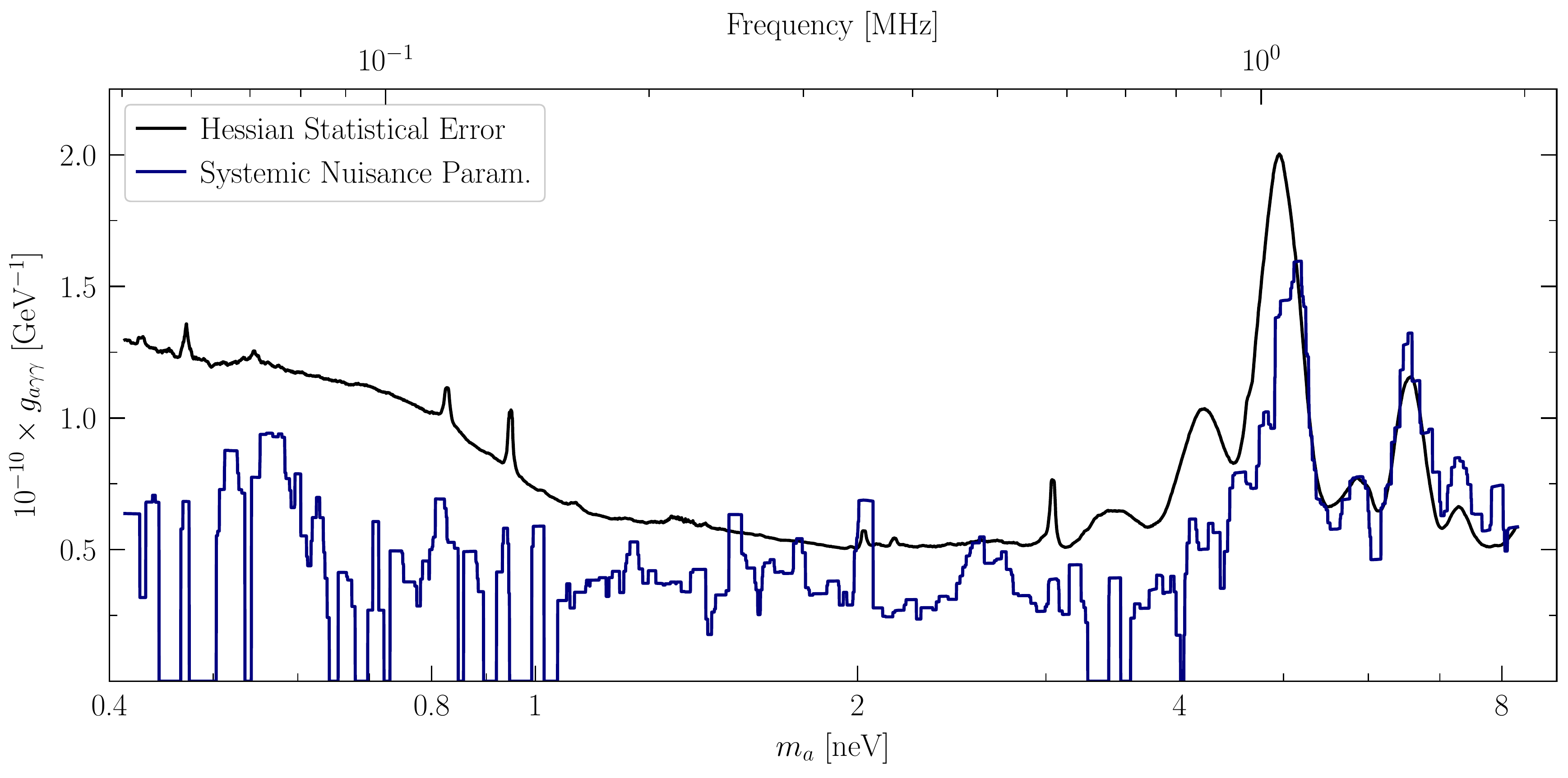}
    \caption{The hyperparameter, $\sigma_A$, converted to the units of $\gagg$, for the systematic nuisance parameter, $g_{a \gamma\gamma}^\mathrm{nuis}$, as a function of axion mass (labeled Systematic Nuisance Param.). We compare the systematic nuisance hyperparameter to the statistical uncertainties (labeled Hessian Statistical Error), which are computed from the Hessian for the log-likelihood without systematic uncertainties about the best-fit axion coupling, $\hat g_{a\gamma\gamma}$. }
    \label{fig:RunSystematic}
\end{figure*}

By construction, the additional nuisance parameter is -- up to a penalty factor -- fully degenerate with the signal.
This allows the background model the flexibility to fit signal-like excesses, but at the cost of a Gaussian penalty factor given by $\mathcal{N}(A_m|0,\sigma_{A_m})$, which is a zero mean normal distribution of width $\sigma_{A_m}$ evaluated at $A_m$.
The magnitude of this penalty is controlled by the hyperparameter $\sigma_{A_m}$, which can be chosen to ensure the above \TS has the expected asymptotic distribution.
To be specific, we determine $\sigma_{A_m}$ for each mass (indexed by $k$) by tuning the observed distribution $\mathrm{TS}(m_a | \sigma_{A_m})$ against the expected distribution in the vicinity of the mass point of interest.
We consider the ensemble of the discovery test statistics belong to the nearest 94,723 mass points, not including: the mass point of interest; the five nearest mass points above and below the mass point of interest; or any mass points that are vetoed by comparison with the magnet off data. 
We then tune the value of $\sigma_{A_m}$ to its minimum value such that there are only three discovery test statistics in excess of 16 within the ensemble, which would be expected if the discovery test statistics were half-chi-square distributed. The nuisance hyperparameter $\sigma_{A_m}$ translated into an effective nuisance parameter $g_{a \gamma \gamma}^\mathrm{nuis}$ is presented in Fig.~\ref{fig:RunSystematic}, and can be understood as an effective floor for our limit-setting power that competes with the statistical noise floor set by the background strength.

We note that many of the procedures required to fix the hyperparameter can be performed analytically.
As the log-likelihood is approximately quadratic around its maximum, $\hat{A}$, near the maximum we have
\begin{equation}
t(m_a, A) = \mathrm{TS}(m_a) \left[1-\left(\frac{A - \hat A}{\hat A} \right)^2\right],
\end{equation}
where at this stage we do not yet zero out test statistics associated with negative best-fit signal amplitudes.
When including the correcting nuisance parameter, the distribution becomes 
\begin{equation}\begin{split}
&t(m_a, A, A_m | \sigma_{A_m})  \\
= &\mathrm{TS}(m_a)\left[1- \left(\frac{A+A_m - \hat A}{\hat A} \right)^2 \right]
- \left(\frac{A_m}{\sigma_{A_m}} \right)^2,
\end{split}\end{equation}
with the final term arising from the Gaussian penalty.
We can now define a new test statistic for discovery including the background signal nuisance parameter as
\es{eq:NuisanceDiscoveryTS}{
    \mathrm{TS}(m_a | \sigma_{A_m}) = &\mathrm{max}_{A, A_m}t(m_a, A, A_n | \sigma_{A_m}) \\
    -&  \mathrm{max}_{A_m} t(m_a, A = 0, A_n | \sigma_{A_m}).
}
Using this, for a given $\sigma_{A_m}$, the new test statistic for discovery with the nuisance background signal can be directly constructed from the test statistic without the nuisance background signal.
In particular, since \eqref{eq:NuisanceDiscoveryTS} involves only maximizations of a quadratic function, the result is given by
\begin{equation}
\mathrm{TS}(m_a | \sigma_{A_n}) =  \frac{\mathrm{TS}(m_a) \hat A^2 }{\hat A^2 + \mathrm{TS}(m_a) \sigma_{A_n}^2},
\end{equation}
which has the effect of decreasing the computed TS.
As before, we then zero out $\mathrm{TS}(m_a | \sigma_{A_n})$ when the best fit signal strength parameter $\hat A$ is negative.

\subsection{Survival Functions, Unvetoed Excesses, and Limits}

The analysis procedure was tuned on 10\% of the Run~2 data.
Once validated, to the full Run~3 dataset, which had remained blinded.
The survival function evaluated at various stages of our analysis procedure, realized on the 10\% of Run~2 data used for tuning, is shown in Fig.~\ref{fig:TestingSurvivals}.
Approximately 10\% of masses in Run~2 and 5\% of masses in Run~3 are removed by the peak exclusion and vetoing procedure, with the fraction of masses removed as a function of frequency shown in Fig.~\ref{fig:acceptance}.

Even after the nuisance parameter tuning, there remain some discrepancies between the observed and expected survival functions at moderate values ($\mathrm{TS} > 16$) of the test statistic.
In particular, there are a small number of mass points which have $\mathrm{TS}$ in excess of our $5\sigma$ LEE threshold in Run~2 data.
All of these excesses occur in nearby frequencies, associated with a transient, and relatively broad, spectral feature which is shown in Fig.~\ref{fig:candidates}.
Further, the mass points which are high significance excesses in the Run~2 data are not significant in the Run~3 data.
Accordingly, we do not consider these excesses to represent credible detections.

\begin{figure*}
    \centering
    \includegraphics[width=\textwidth]{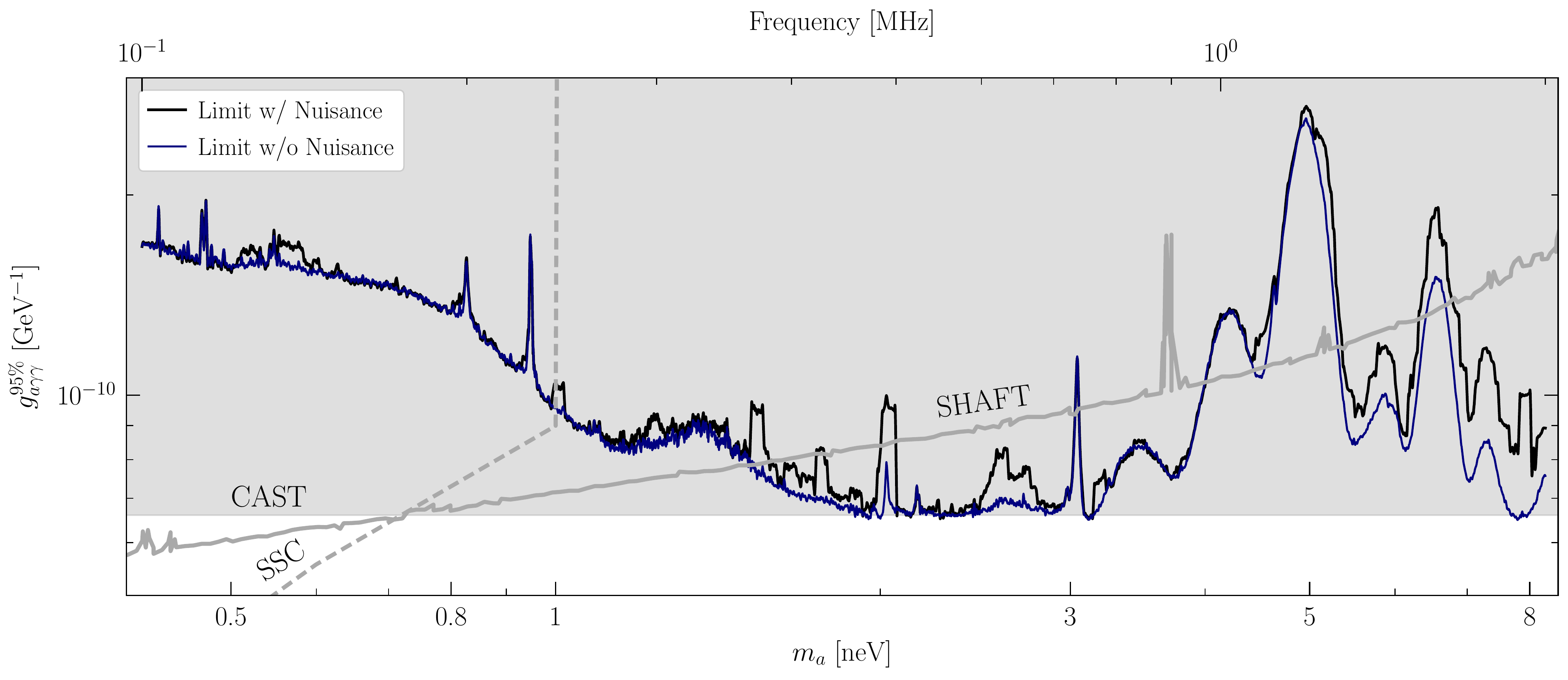}
    \caption{A comparison of our fiducial limits that include a nuisance hyperparameter correction (black) and those without any correction (blue). Limits set with the nuisance hyperparameter are slightly weaker, but the features and limit-setting power are broadly similar. The figure is smoothed for clarity.}
    \label{fig:RunLimits_NN}
\end{figure*}

The independent Run~3 limits are shown with and without the tuned nuisance parameter in Fig.~\ref{fig:RunLimits_NN}.

\subsection{Injected Signal Tests}

To further validate the robustness of our analysis framework, we can inject a synthetic signal into the data and confirm that: (1) we are able to recover the signal strength, when expected; and (2) our limits will not exclude an injected signal.
To perform this test, we select five representative mass points and inspect the real data in the vicinity of the expected location of an injected signal.
We generate independently drawn axion signals at a range of axion couplings strengths which we add on top of each of the spectra collected in Run~3.
We then apply our analysis framework, adopting the tuned value of the nuisance parameter that was previously determined from the real data in the vicinity of the injected signal location, and evaluate the best-fit axion coupling, the $95^\mathrm{th}$ percentile upper limit on that coupling, and the detection significance as a function of the true axion coupling of the injected signal.
As a further test of the performance of our analysis framework, for each of the five mass points, we fit the Run~3 data under the null model.
We then generate Monte Carlo data under null model fits and repeat our procedure of injecting and analyzing, allowing us to compare the analysis of signal injections on real data with expected performance of the analysis framework under the null model.
With the exception of the tuned nuisance parameter, which we continue to keep fixed at its value determined from the real data, this represents an entirely self-contained test of the analysis procedure.

\begin{figure*}
    \centering
    \includegraphics[width=.95\textwidth]{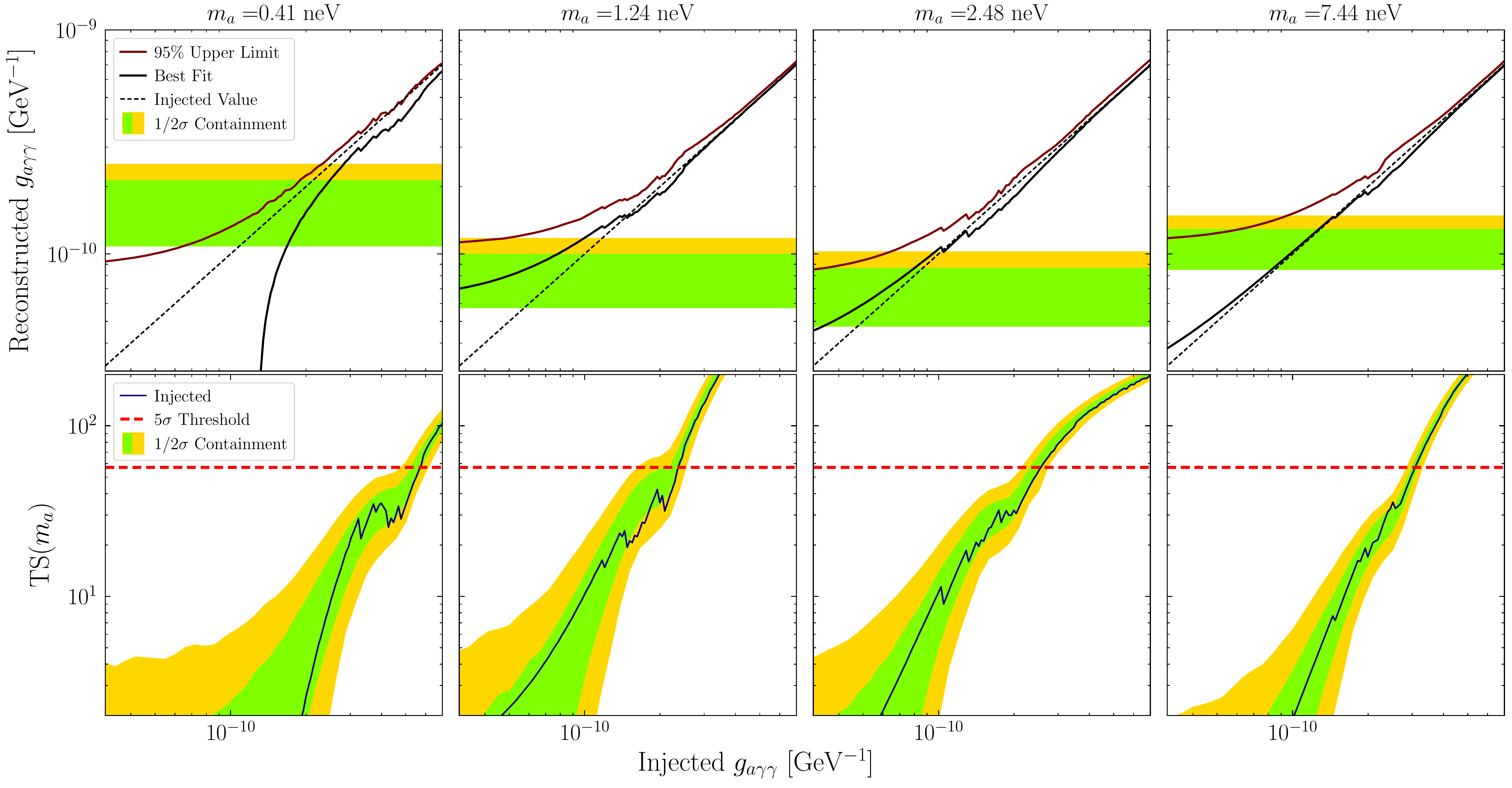}
	\caption{(\textit{Top row}) The best fit and 95\% upper limit on the recovered signal strength as a function of the injected signal strength at five mass points evaluated on the real Run~3 data. The results are compared to the $1\sigma$ and $2\sigma$ expectations for the $95^\mathrm{th}$ percentile upper limit under the hypothesis of no axion signal as determined by 2560 Monte Carlo (MC) realizations of the null model fits to the real data at each injected signal strength. (\textit{Bottom row}) In black, the recovered detection test statistic for the signal injected in the real data as a function of injected signal strength. The dashed red line indicates the threshold for a $5\sigma$ detection significance account for the look-elsewhere effect while the green and yellow bands indicate the $1\sigma$ and $2\sigma$ expectations for the detection significance determined from 2560 MC realizations of the null model combined with appropriately varied injected signal strength.}
    \label{fig:SignalInjection}
\end{figure*}

The results of these tests are shown in  Fig.~\ref{fig:SignalInjection}.
Critically, our analysis procedure is able to place a robust 95$^\mathrm{th}$ percentile upper limit which does not exclude the true coupling strength at which the signal is injected more often than would be expected and accurately recovers the correct axion parameters at a detection significance within the simulated expectations. We also briefly comment on the somewhat jagged nature of the detection significance in the real data as a function of the injected signal strength. These features are a product of the filtering included in our analysis procedure which removes at most 50\% of the spectra in $\sim$5\% subintervals if those subintervals have a detection significance in excess of $3\sigma$. This has the effect of somewhat weakening the detection significance in discretized steps and also slightly biases the 95$^\mathrm{th}$ percentile limit to slightly lower values. This bias is removed using a TS-dependent correction of at most 8\% that is incorporated in our limits.

\begin{figure*}
    \centering
    \includegraphics[width=\textwidth]{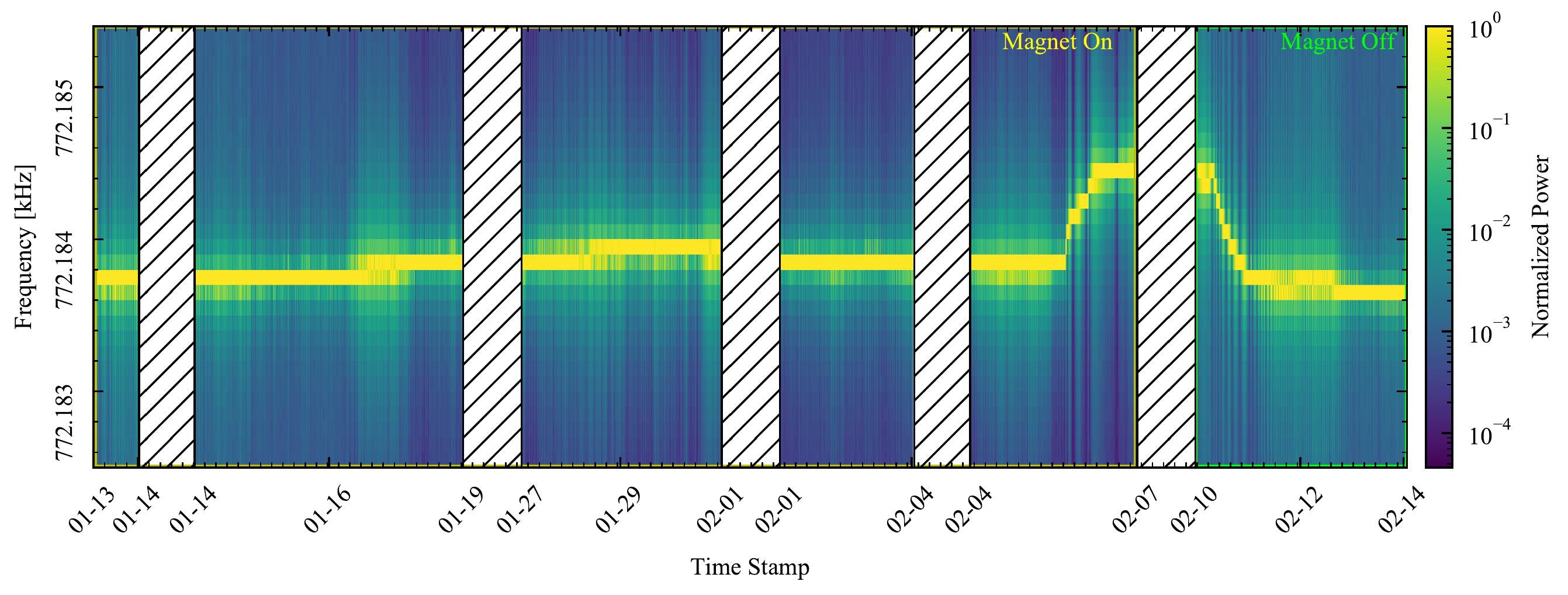}
    \caption{The time evolution of the broad excess that is associated with the putative signal candidate in the Run~2 data that survived all analysis cuts. The excess persists after the magnet is turned off and evolves in frequency, indicative of a background source.  The magnet off veto did not anticipate this level of time evolution and so did not remove these excesses. Since this feature was found after unblinding, we report it here but do not consider it to be a credible axion detection.}
    \label{fig:candidates}
\end{figure*}

\end{document}